\documentclass[twocolumn,showpacs,preprintnumbers,amsmath,amssymb,superscriptaddress]{revtex4-1}
\usepackage[dvipdfm]{graphicx}
\usepackage{aas_macros}

\usepackage{ulem}
\usepackage{color}
\input{colordvi.tex}

\begin{document}
\title{Exploring the origin of the fine structures in the CMB temperature angular power spectrum}
\author{Kohei Kumazaki}
\email{kumazaki@a.phys.nagoya-u.ac.jp}
\affiliation{Department of Physics and Astrophysics, Nagoya University, Nagoya 464-8602, Japan}

\author{Kiyotomo Ichiki}
\affiliation{Department of Physics and Astrophysics, Nagoya University, Nagoya 464-8602, Japan}

\author{Naoshi Sugiyama}
\affiliation{Department of Physics and Astrophysics, Nagoya University, Nagoya 464-8602, Japan}
\affiliation{Institute for the Physics and Mathematics of the Universe (IPMU), The University of 
  Tokyo, Chiba 277-8582, Japan}
\affiliation{Kobayashi-Maskawa Institute for the Origin of Particles and the Universe, 
  Nagoya University, Nagoya 464-8602, Japan}

\author{Joseph Silk}
\affiliation{Department of Physics, University of Oxford, Keble Road, Oxford, OXI 3RH, UK}

\date{\today}
\begin{abstract}
  The angular power spectrum of the cosmic microwave background (CMB)
  temperature anisotropies is a good probe to look into the primordial
  density fluctuations at large scales in the universe.  Here we
  re-examine the angular power spectrum of the Wilkinson Microwave
  Anisotropy Probe data, paying particular attention to the fine
  structures (oscillations) at $\ell=100 \sim 150$ reported by several
  authors.  Using Monte-Carlo simulations, we confirm that the gap from
  the simple power law spectrum is a rare event, about 2.5--3$\sigma$,
  if these fine structures are generated by experimental noise and the
  cosmic variance.  Next, in order to investigate the origin of the
  structures, we examine frequency and direction dependencies of the
  fine structures by dividing the observed QUV frequency maps into four
  sky regions.  We find that the structures around $\ell \sim 120$ do
  not have significant dependences either on frequencies or directions.
  For the structure around $\ell \sim 140$, however, we find that the
 characteristic signature found in the all sky power spectrum is
 attributed to the anomaly only  in the South East region.  
\end{abstract}
\maketitle

\section{Introduction}

The inflationary cosmology is a successful paradigm in explaining the
generation of primordial density fluctuations and solving essential problems of
the classical Big Bang cosmology
\cite{1981PhRvD..23..347G, 1981MNRAS.195..467S, 1980PhLB...91...99S, 1982PhLB..115..295H,1982PhLB..117..175S, 
1982PhRvL..49.1110G}.  The primordial density fluctuations
are transformed into the anisotropy of the Cosmic Microwave Background
(CMB) and the Large Scale Structure of the universe (LSS).  Thanks to
the high angular resolution and longer-term observations of the
anisotropy of the temperature fluctuations, such as by the Wilkinson
Microwave Anisotropy Probe (WMAP) \cite{Jarosik:WMAP7yr}, the South Pole
Telescope \cite{2011ApJ...743...28K}, the Atacama Cosmology Telescope
\cite{2011ApJ...739...52D}, the Arcminute Cosmology Bolometer Array Receiver
\cite{2009ApJ...694.1200R}, the Cosmic Background Imager
\cite{2009arXiv0901.4540S} and so on,  it has been found that the angular 
power spectrum of temperature fluctuations conforms to the prediction
from the $\Lambda$CDM model with slow roll inflation.

While the observed angular power spectrum is globally consistent with a
smooth power-law primordial spectrum of density fluctuations 
\cite{Wang:1998gb, Tegmark:2002cy, Spergel:2006hy, Vazquez:2011xa, Hlozek:2011pc,}, 
some gaps
between the prediction from the simplest power-law model and the
observed data have been reported. They are discussed recently by grace
of the detailed observations, including a small bump and dip at
$\ell = 20$ -- $40$ \cite{Mortonson:2009qv,Dvorkin:2009ne,Hazra:2010ve} or
oscillation around $\ell = 100$ -- $150$
\cite{Ichiki:2009zz,Ichiki:2009xs,Nakashima:2010sa,Kumazaki:2011eb}.

Possible causes of these anomalous structures are discussed in the
literature.  The small bump and dip at $\ell=20$ -- $40$, for example,
may be explained by the mass variation of the inflaton during the
inflation phase
\cite{Adams:2001vc,Mortonson:2009qv,Dvorkin:2009ne,Hazra:2010ve}.  When
inflaton obeys single slow-roll inflation dynamics, the power spectrum
of primordial density fluctuations shows a power-law feature, and the
angular power spectrum of the CMB is expected to be a smooth curve.  If the
inflaton mass has changed during inflation, however, some oscillating
structures emerge in the power-law primordial power spectrum because the
inflaton field is forced to accelerate and/or decelerate rapidly during
the slow roll inflation phase.  In such cases the bump and dip structure
rises up in the angular power spectrum of the CMB.

On the other hand, it seems difficult to explain the oscillating
structures around $\ell = 100$ -- $150$ on the firm theoretical
background, though some works have tried to explain the origin
\cite{Ichiki:2009zz,Ichiki:2009xs,Nakashima:2010sa,Kumazaki:2011eb}.  In
the previous paper, some of us have tried to explain the
oscillating structures with the inflaton mass variation
\cite{Kumazaki:2011eb}.  The condition considered in that paper is that
inflaton mass changes with oscillations.  We found that oscillating
structures can be generated at the arbitrary scale by adjusting
oscillation number and time scale of the mass variation.  However, the
width of this structure tends to become so wide, and to match the
observed data we need some fine-tunings.  If we force to explain the
oscillation matching with the observed data, the parameters become
unrealistic values. Therefore, we have concluded that it is difficult to
explain the oscillating structure on the angular power spectrum found at
multipole range $\ell=100$--150 with the inflaton mass variation.

Nakashima et al. \cite{Nakashima:2010sa} have proposed a sudden change
of the sound velocity of the inflaton field during inflation.  Based on 
their model, the authors found oscillating structures in the primordial
power spectrum. However, the oscillating structures tend to extend in a wide
range of wavenumber up to $\ell \leq 300$ which includes the first acoustic
peak.  Therefore, this model may not match with observed data which
shows the oscillation only at the confined region of $100\leq \ell\leq 150$.

In light of the difficulty in explaining the structure on the
theoretical background, in this paper we closely explore the origin of
the structure with the data taken by WMAP, particularly paying attention
to the frequency and direction dependences. If the observed structure is
really from the cosmological origin, we expect the structures should be
independent of them.

This paper is organized as follows.  In section II, we review the method
to estimate the angular power spectrum from the two point correlation
function.  Following the method, we examine the probability of the fine
structures being generated from noise and/or cosmic variance in section
III, and estimate the angular power spectrum with each frequency band in
section IV.  In section V, we give the angular power spectrum with the
partial sky, with an explanation of some difficulties in estimating the
angular power spectrum with the partial sky map
\cite{Ansari:2009ys,Ko:2011ut}.  Section VI is devoted to the summary of
this work.

\section{The angular power spectrum}
The angular power spectrum $C_\ell$ is a good indicator for quantifying
the temperature fluctuations of the CMB, and defined as
\begin{equation}
  \langle\,a^\ast_{\ell m}a_{\ell^\prime m^\prime}\,\rangle 
  = \delta_{\ell\ell^\prime}\delta_{m m^\prime}C_\ell~,
\end{equation}
with 
\begin{equation}
  a_{\ell m} = \int \sin{\theta} d\theta \int d\phi 
  \Delta T(\theta, \phi)Y_{\ell m}^\ast(\theta, \phi)\,,
\end{equation}
where $Y_{\ell m}^\ast(\theta, \phi)$ is the spherical harmonics
function evaluated at the position $(\theta, \phi)$ of spherical
coordinate,  $\Delta T(\theta, \phi)$ is the value of temperature
fluctuations and $\langle$...$ \rangle$ means
ensemble average.
In practice, however, we can observe only one unique sky and need to estimate
the angular power spectrum from one realization. 
We can obtain the estimator $\tilde C_\ell$ of the $C_\ell$ by
\begin{equation}
\tilde C_\ell = \frac{1}{2\ell +1}\sum^\ell_{\ell=-m}|a_{\ell m}|^2.  
\end{equation}
In this work, we use a fast CMB analysis named
Spice introduced by Szapudi et al. for temperature \cite{Szapudi:2000xj}
and Chon et al. for polarization \cite{2004MNRAS.350..914C}.  
This software calculates the angular power spectrum from the CMB 
temperature/polarization data and can convert the angular power spectrum 
to two point correlation function or vice versa, if we need.  
We review the relation between the angular power spectrum and 
the angular correlation function for later discussion.  

The CMB temperature angular correlation function $\xi(\theta)$ is defined by
\begin{equation}
  \xi(\theta) = \langle\,\Delta T(\vec n) \Delta T(\vec n^\prime)\,\rangle~, 
\end{equation}
where $\vec n$ and $\vec n^\prime$ denote sky directions, and $\cos
\theta = \vec n \cdot \vec n^\prime$. 
The relation between the angular correlation function
and the angular power spectrum can be written as,
\begin{eqnarray}
  \xi(\theta) = \frac{1}{4\pi}\sum_{\ell=0}^{\infty}\,(2\ell+1)\,C_\ell\,P_\ell(\cos\theta)~, \\
  C_\ell = 2\pi \int_{0}^\pi \sin\theta\,d\theta\,\xi(\theta)\,P_\ell(\cos\theta)~, 
\end{eqnarray}
where $P_\ell(\cos\theta)$ is the Legendre polynomial.

%%%%%%%%%%%%%%%%%%%%%%%%%%%%%%%%%%%%%%%%%%%%%%%%
%%%
%%% Validity of the fine structure
%%%
%%%%%%%%%%%%%%%%%%%%%%%%%%%%%%%%%%%%%%%%%%%%%%%%

\section{Validity of the fine structures}
\label{sec:vldty} 
\subsection{Monte-Carlo method}
As a possible origin of the fine structures,
instrumental noise and cosmic variance effects should be called into
question first.  In this section, we examine the possibility of the fine
structures being generated by these noise effects.  For this purpose, we
generate a number of sky maps of CMB temperature fluctuations using the
HEALPix function, {\bf isynfast} \cite{Gorski:1998vw}.  The routine
generates a temperature fluctuation map from an input angular power
spectrum $C_\ell$.  For the $C_\ell$, we adopt the best-fitting
$\Lambda$CDM model of the WMAP seven year parameter table
\cite{Komatsu:WMAP7yr}.  The resolution of this map is taken as $N_{\rm
side}=512$.  This value is same as that of the released sky map given by
the WMAP team.  

Next, we add the Gaussian noise expected for the WMAP observation on
these simulated sky maps.  The variance of the instrumental noise can be written by
\begin{equation}
  N^i=\frac{\sigma_0}{\sqrt{N_{\rm obs}^i}}~, 
\end{equation}
where superscript $i$ represents the pixel number on the sky, $\sigma_0$
the rms noise per an observation, and $N_{\rm obs}^i$ the number of
observation of $i$th pixel.  The values of $\sigma_0$ and $N^i_{\rm
obs}$ are also given by the WMAP team \cite{Jarosik:WMAP7yr}.  In this
way, we can prepare the sky maps of CMB temperature fluctuations of
different universes with the same cosmological parameters and noise
properties.

\begin{figure}[t]
  \begin{center}
    \includegraphics[width=0.45\textwidth] {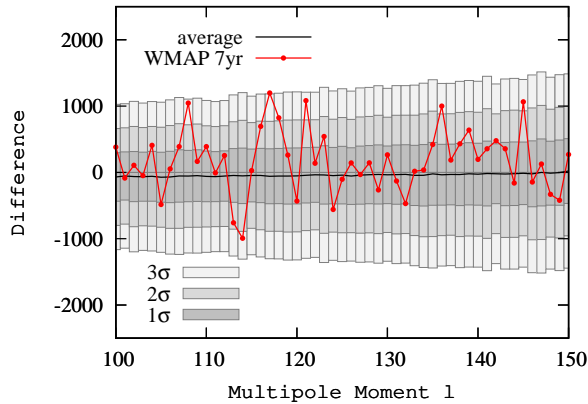}
    \caption{Differences between the WMAP angular power spectrum
    $C_\ell^{\rm WMAP}$ and the power spectrum of the $\Lambda$CDM model
    $C_\ell^{\rm \Lambda CDM}$ (red line). The boxes represent the
    variance of $1,2,3\sigma$ at each $\ell$ estimated from the 3,0000
    simulations.  The black line represents the average value of
    residuals between $C_\ell^j$'s and $C_\ell^{\rm \Lambda CDM}$.}
    \label{fig:val}
  \end{center}
\end{figure}

In the present analysis, we prepare 3,000 sky maps and calculate the
angular power spectrum for each sky map.  These $C_\ell$'s are denoted by
$C_\ell^j$'s using a superscript of the map number $j$.  We estimate the
average and variance of these $C_\ell^j$'s at each multipole moment
$\ell$, and compare the angular power spectra with the WMAP angular
power spectrum, $C_\ell^{\rm WMAP}$.  In order to emphasize the
differences from the $\Lambda$CDM model, we show the distributions of
the simulated $C_{\ell}$'s as residuals from the $\Lambda$CDM model,
$C_{\ell}^{\rm \Lambda CDM}$ in Fig.~\ref{fig:val}.  The variances of
$C_\ell$'s at $1\sigma$, $2\sigma$, and $3\sigma$ are shown as the boxes.
We can see that the average of the simulated $C_\ell$'s (the black line
in the figure) is on the zero horizontal axis.  The red line represents
the residuals between $C_\ell^{\rm \Lambda CDM}$ and $C_\ell^{\rm
WMAP}$.

\subsection{significance of the structure}
In the multipole region of $\ell = 100$ -- $150$, the distributions of
the $C_\ell$'s are well approximated by the Gaussian one. In that case
the probability that the $C_\ell$ is within $1\sigma$ and $2\sigma$
should correspond to about 68.3\% and 95.5\%, respectively.  In our
analysis, there are 51 data points because we focus only between
$\ell=100$ -- $150$, and the expected number of data over $1\sigma$
or $2\sigma$ is estimated as 16.2 or 2.3 points, respectively.
Thus, it is natural that some data points deviate from the theoretical
curve to this extent.  However, from Fig.~\ref{fig:val}, we find that 19
and 7 points of data are over
$1\sigma$ and $2\sigma$ levels, respectively.  That indicates the
$C_\ell^{\rm WMAP}$ somewhat deviates from the standard Gaussian
distribution more than expected.

\begin{figure}[thbp]
%  \centering \includegraphics[width=0.5\textwidth, bb= 70 40 400 300]
   \centering \includegraphics[width=0.5\textwidth]
  {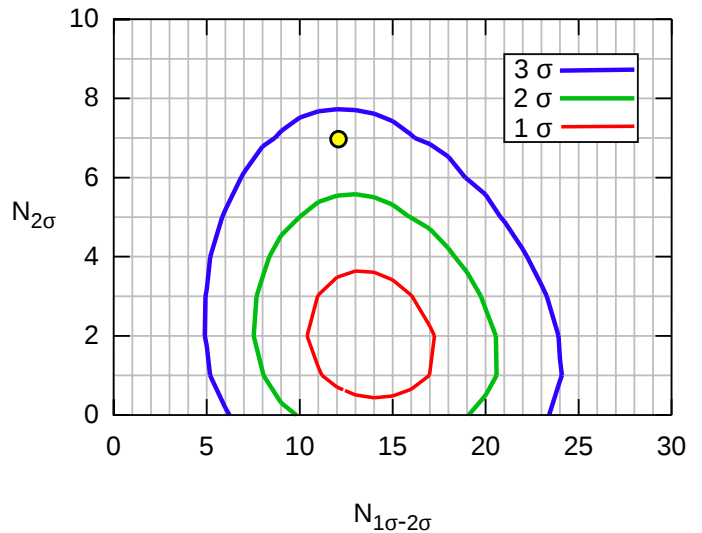} \caption{Contours of equal probabilities as a
  function of $N_{1\sigma\mbox{-}2\sigma}$ and $N_{2\sigma}$.  The red,
  green and blue lines correspond to $1\sigma$, $2\sigma$ and $3\sigma$,
  respectively. The (yellow) point represents the WMAP data.}
  \label{fig:possibility}
\end{figure}

We roughly estimate the probability $P$ as a function of $N_{2\sigma}$
and $N_{1\sigma\mbox{-}2\sigma}$ which are the numbers of realization
that exceeds $2\sigma$ and lies between 1$\sigma$ and 2$\sigma$,
respectively, assuming the realization follows the Gaussian
distribution.  We show the result as contours in
Figure.~\ref{fig:possibility}.  In the figure the red, green and blue
lines represent $1\sigma$, $2\sigma$ and $3\sigma$, respectively.  From
the figure, we can see that the peak is at the expected value
($N_{1\sigma\mbox{-}2\sigma}$,\, $N_{2\sigma}$)\,=\,(13.9,\, 2.3), and
the probability decreases with distance away from the expected value.
The yellow point represents the value from the WMAP data;
($N_{1\sigma\mbox{-}2\sigma}$,\, $N_{2\sigma}$)\,=\,(12,\, 7).  It lies
at the location between 2.5--$3\sigma$, which is consistent with the
result of \cite{Ichiki:2009zz}.  Therefore, we can understand
that the fine structures at the range of $\ell=100$--150 of the angular
power spectrum which are observed by WMAP team seem a rare event, if
these structures are generated only by the noise and the cosmic variance
effects.

If the structures were not related to the noise nor cosmic variance, the
other possibilities would be foreground effects and/or real cosmological
signal. From the next section, we examine frequency and direction
dependences of the fine structures in order to investigate whether the
structures come from cosmological or astronomical phenomena.

%%%%%%%%%%%%%%%%%%%%%%%%%%%%%%%%%%%%%%%%%%%%%%%%
%%%
%%% Frequency dependence
%%%
%%%%%%%%%%%%%%%%%%%%%%%%%%%%%%%%%%%%%%%%%%%%%%%%

\section{Frequency dependence}
If any astronomical phenomena such as syncrotron emission, dust
emission and radio galaxies infiltrate the CMB temperature
fluctuation data and create the fine structures, we expect that they
have a characteristic frequency dependence.  Therefore, we estimate the
angular power spectra for three different frequency bands, namely Q, V
and W bands.  We find that the shapes of the fine structures at each
frequency band are very similar to each other and to the all sky one
(Fig.~\ref{fig:val}).  Thus it is improbable that
the fine structures originate from astrophysical phenomena nor objects,
because they will have different frequency dependences from the CMB
blackbody spectrum.

%%%%%%%%%%%%%%%%%%%%%%%%%%%%%%%%%%%%%%%%%%%%%%%%
%%%
%%% Direction dependence
%%%
%%%%%%%%%%%%%%%%%%%%%%%%%%%%%%%%%%%%%%%%%%%%%%%%

\section{Analysis with the partial skies}

\begin{figure}
  \begin{tabular}{cc}
    \begin{minipage}{0.25\textwidth}
      \begin{center}
        \includegraphics[width=\textwidth]
                        {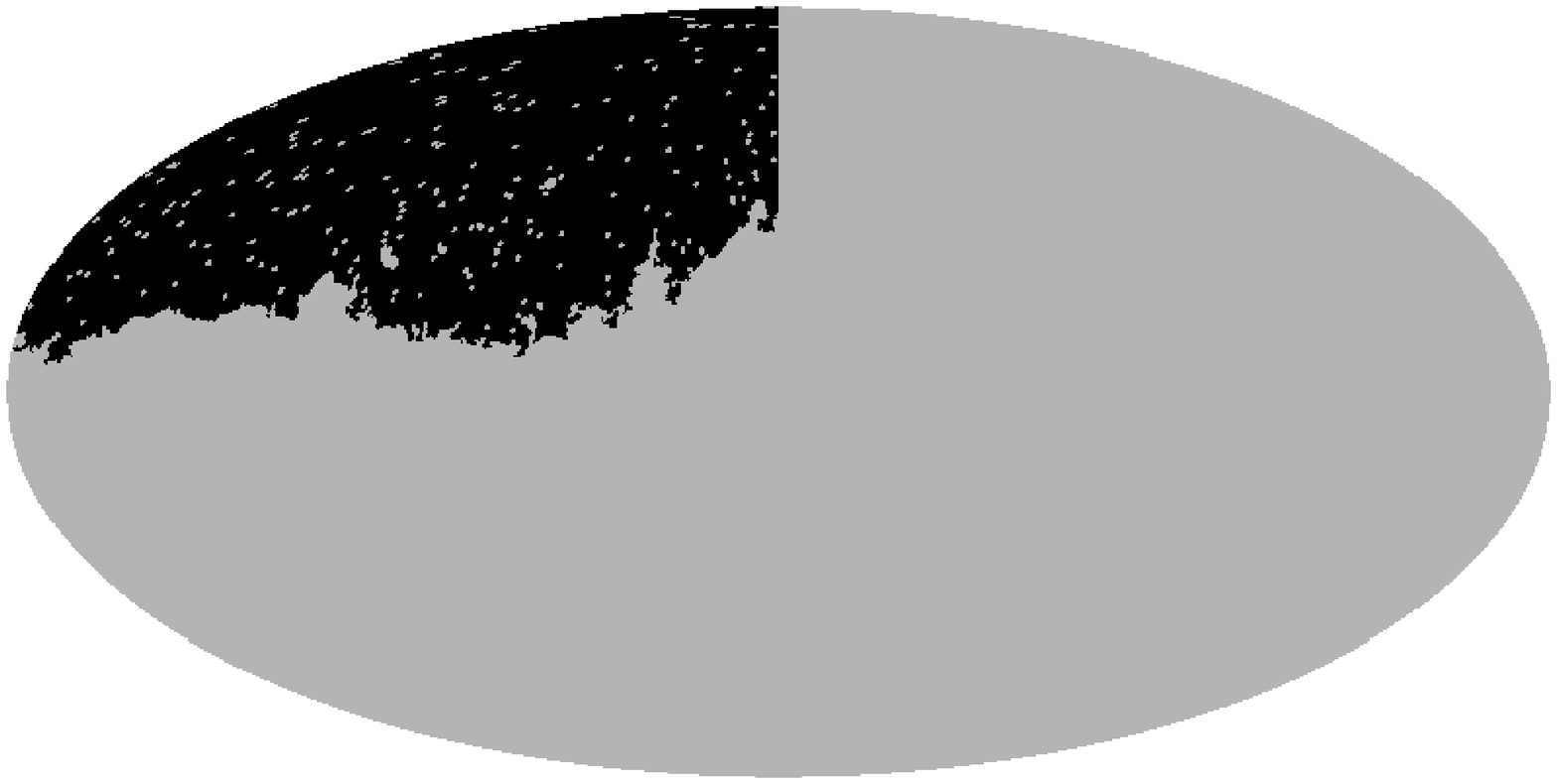}
      \end{center}
    \end{minipage}
    \begin{minipage}{0.25\textwidth}
      \begin{center}
        \includegraphics[width=\textwidth]
                        {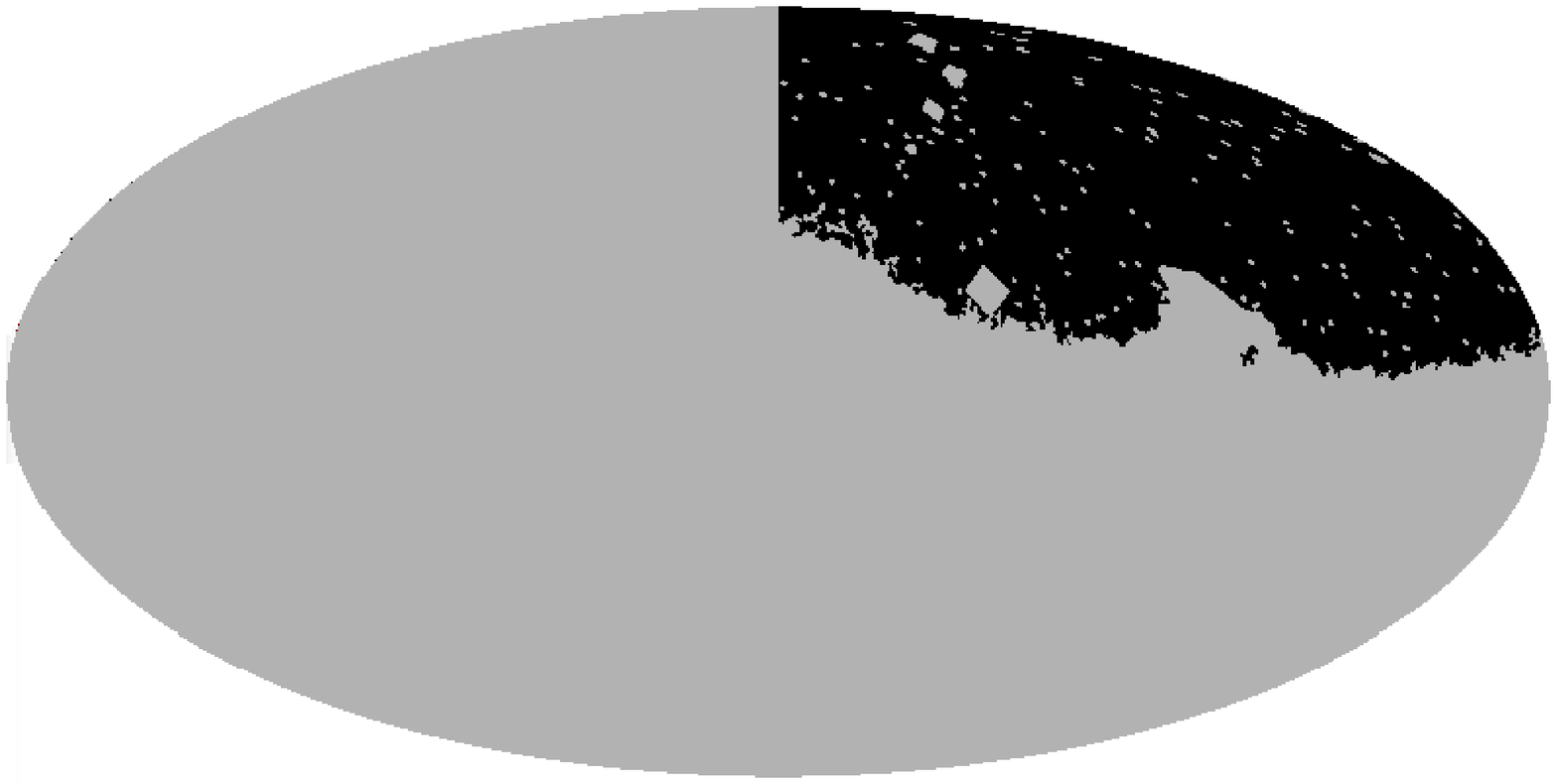}
      \end{center}
    \end{minipage}
  \end{tabular}
  \begin{tabular}{cc}
    \begin{minipage}{0.25\textwidth}
      \begin{center}
        \includegraphics[width=\textwidth]
                        {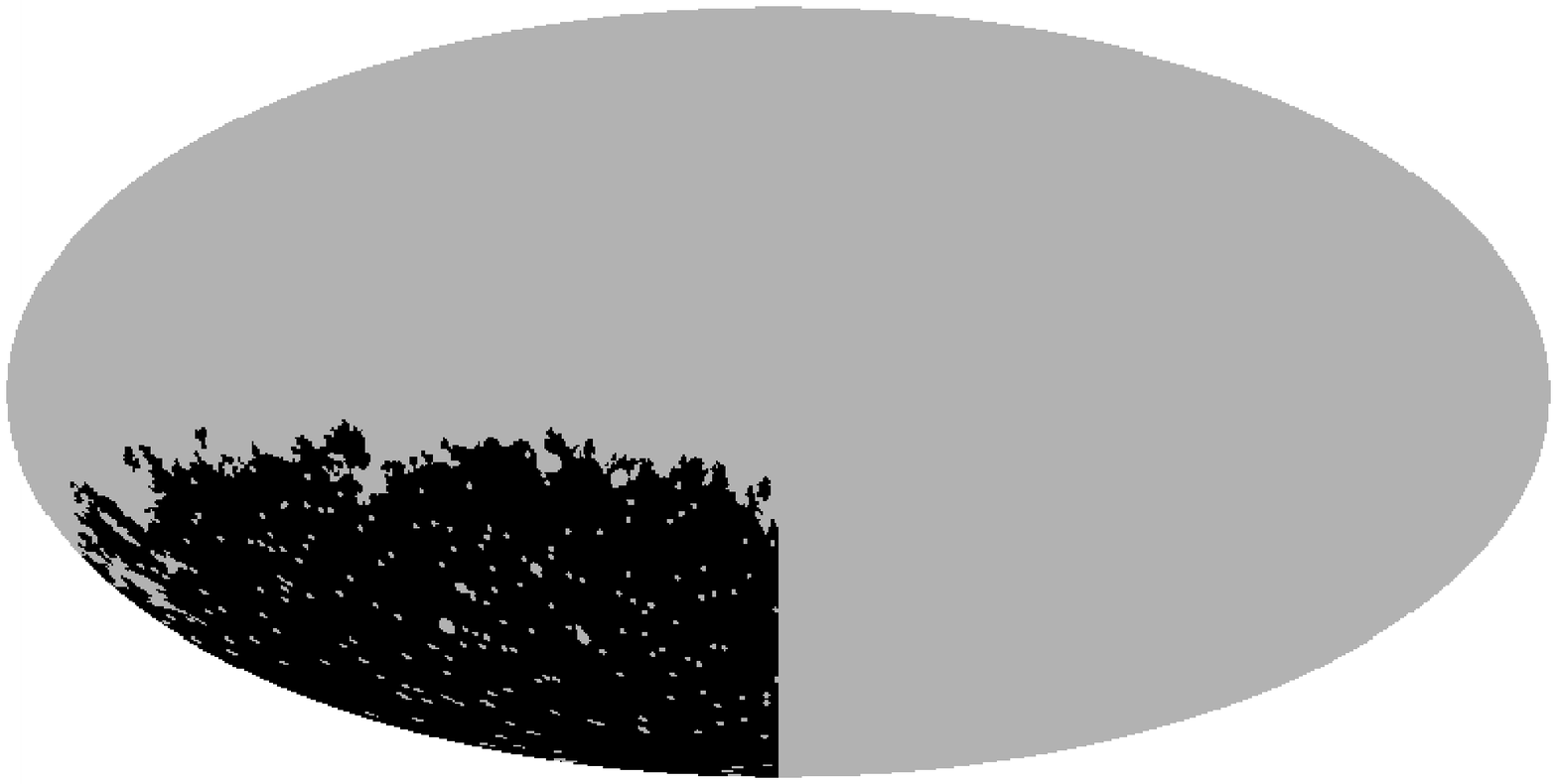}
      \end{center}
    \end{minipage}
    \begin{minipage}{0.25\textwidth}
      \begin{center}
        \includegraphics[width=\textwidth]
                        {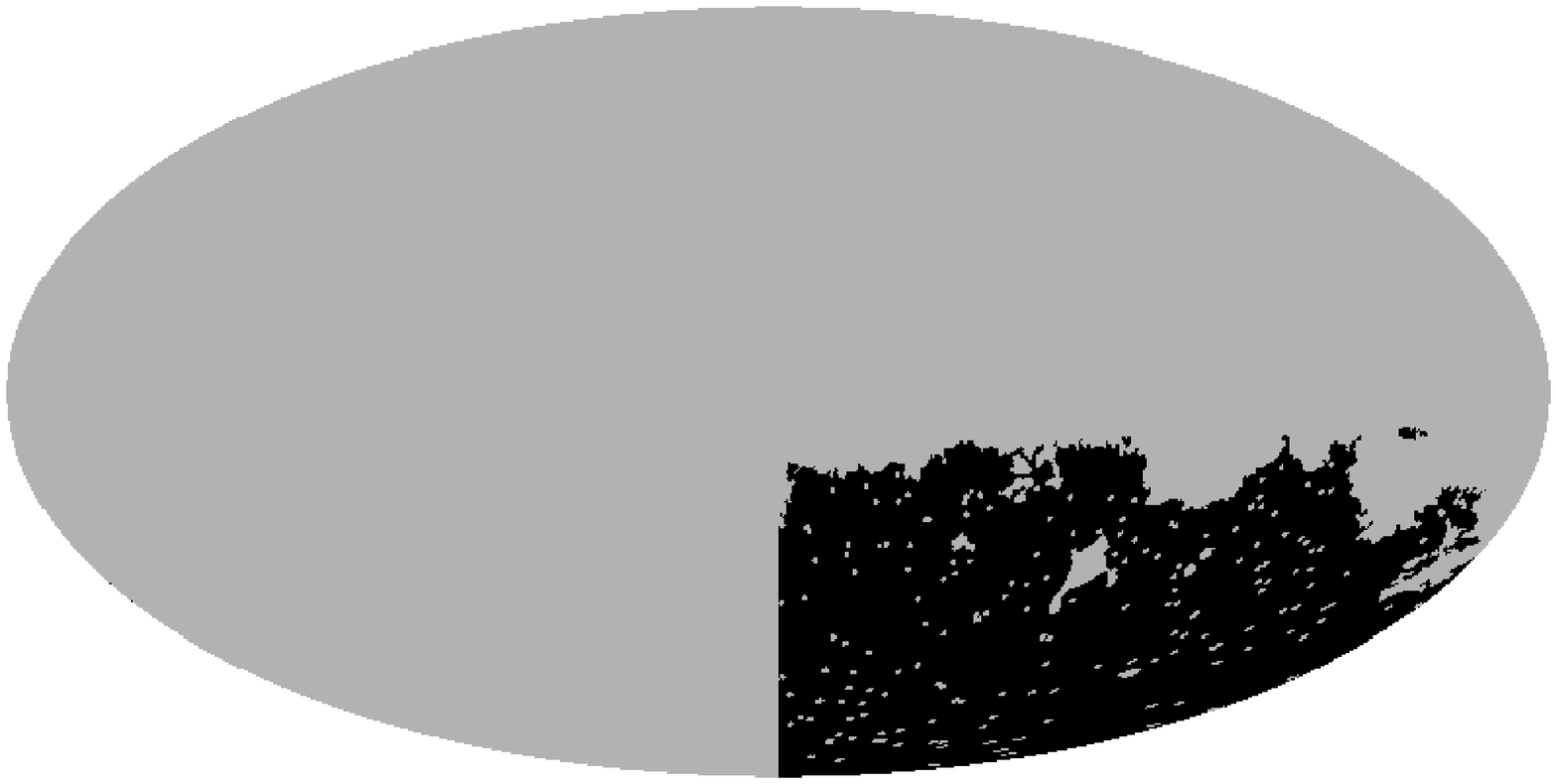}
      \end{center}
    \end{minipage}
  \end{tabular}
  \caption{Masks of the partial skies used in our analysis in the
 galactic coordinate. 
 The left top panel refers to North West mask, the right top panel to North East mask, 
 the left bottom panel to South West mask and, finally, the right bottom to South East mask.}
  \label{fig:mask_maps}
\end{figure}

In this section, we look into the differences in the angular power
spectra for different sky directions.  For this purpose we prepare
four masks, namely the North West mask ($0\leq \phi\leq \pi$, $0\leq
\theta\leq \pi/2$), the North East mask ($\pi\leq \phi\leq 2\pi,\,0\leq
\theta\leq \pi/2$), the South West mask ($0\leq \phi\leq \pi,\,\pi/2\leq
\theta\leq \pi$) and the South East mask ($\pi\leq \phi\leq 2\pi,\,\pi/2\leq
\theta\leq \pi$).  We show these masks in Fig.~\ref{fig:mask_maps}.

Using the masks, first let us calculate the gap between the angular
power spectra of the simulated maps and that of the observed map by WMAP
for the partial skies.  Further, we compare the partial sky maps with
each other, in order to see the dependence of the angular power spectrum
on the sky directions.  However, some difficulties arise when analyzing
the CMB map with a partial sky mask.  Because a mask narrows down the
effective area of the spherical surface, effective information is
reduced.  Furthermore, if a mask has a sharp cutoff at the boundary,
it will lead a truncated correlation function. Computing the
angular power spectrum from the truncated correlation function causes
spurious oscillations inherent to the Fourier transformation, which is
known as the "Gibbs phenomenon".  However, these oscillations have
nothing to do with the anisotropy of the CMB nor the fine structures,
and they can be removed by apodization and binning techniques as we
describe below.

In the next two subsections, we review these techniques
that we used in order to lighten and avoid these nuisance effects, and
in the final subsection, we show the direction dependence of the CMB
angular power spectrum.

\subsection{Apodization}
In fact, we find that spurious oscillatory structure in the angular
power spectrum that arises when the masks are used is caused by the
large amplitude oscillations in the two point correlation function
at large angular scales. The oscillations are caused by the
statistical errors. Indeed, if we use a map with a maximal angular size
$\theta_{\rm max}$, nothing can be known about the correlation function
for $\theta \gtrsim \theta_{\rm max}$. However, simply truncating
the correlation function at $\theta=\theta_{\rm max}$ does not help the
situation as we mentioned above.  Instead, the technique uses an appropriate
function, called the "apodization function" $F(\theta)$, and
multiplies it with the correlation function $\xi(\theta)$ for
the product to go to zero smoothly \cite{Szapudi:2000xj,2004MNRAS.350..914C}.  
Angular power spectrum is then given by
\begin{eqnarray}
  C^{\rm apd}_\ell = 
  2\pi\int_{\theta=0}^{\theta_{\rm max}}\,\sin\theta d\theta \,\,
  \xi(\theta)F(\theta)P_\ell(\cos\theta)~,
  \label{eq:7}
\end{eqnarray}
where $\theta_{\rm max}$ is the maximum angle set by the mask.  We can
adopt any function as an apodization function which takes $F(\theta)=1$
at $\theta=0$ and decreases as $\theta$ increases.  The
Gaussian type $F_{\rm G}(\theta)$ and Cosine type $F_{\rm C}(\theta)$
are often adopted, and they are defined as
\begin{eqnarray}
  F_{\rm G}(\theta) &=& \exp\left[-\frac{\theta^2}{2(\sqrt{8\ln 2}\,\sigma_{\rm apd})^2}\right],\\[0.2cm]
  F_{\rm C}(\theta) &=& \frac12\left[1+\cos\left(\pi\frac{\theta}{\sigma_{\rm apd}}\right)\right]~,
\end{eqnarray}
where $\sigma_{\rm apd} = \pi\theta_{\rm apd}/180^\circ$ and $\theta_{\rm apd}$
represents the angle in degree where the apodization
function becomes close to zero.  Typical value of $\theta_{\rm apd}$ should be
close to the cut off angle $\theta_{\rm max}$.

\begin{figure}
    \begin{minipage}{0.5\textwidth}
      \centering
      \includegraphics[width=0.9\textwidth]
      {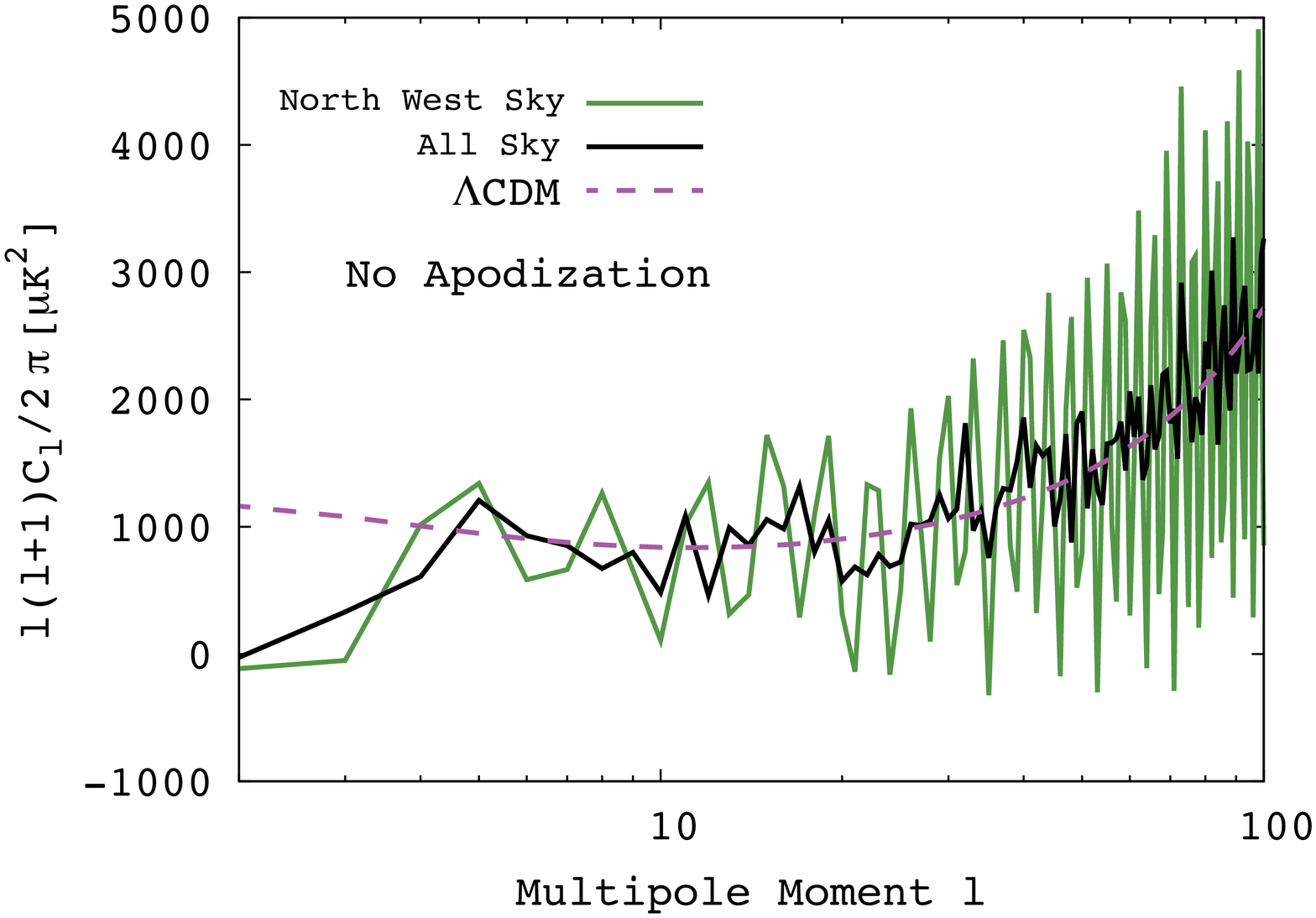}
    \end{minipage}
    \begin{minipage}{0.5\textwidth}
      \centering
        \includegraphics[width=0.9\textwidth]
        {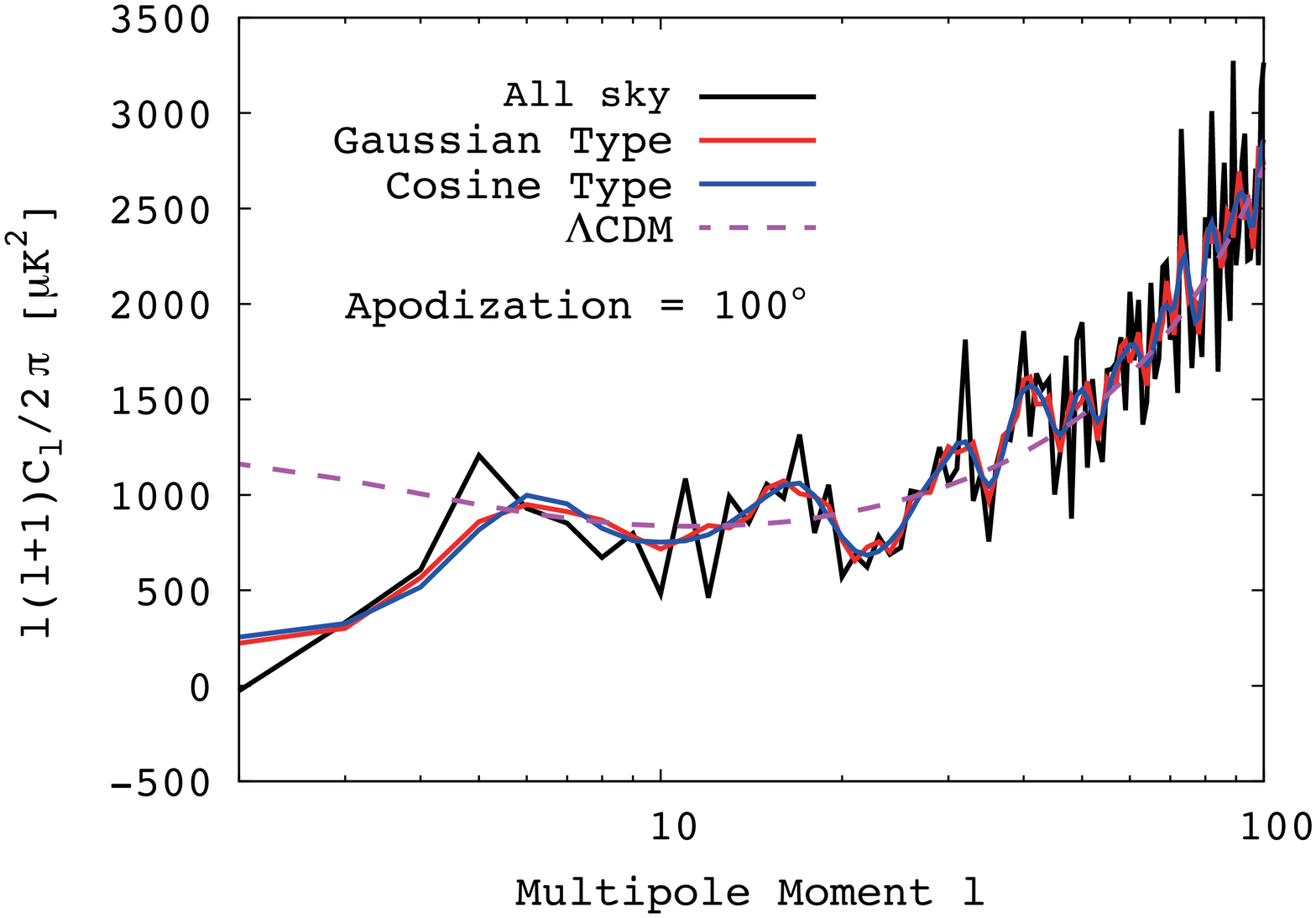}
    \end{minipage}
  \caption{Angular power spectra estimated from two-point correlation
    functions.  In the upper panel, the red line represents the power
    with the North West mask and with $\theta_{\rm max}=100^\circ$ in
    Eq.(\ref{eq:7}), (which corresponds to a truncated correlation
    function) and the black line the case with KQ75 mask.  The
    theoretical curve of the $\Lambda$CDM model is also shown (the
    magenta line).  When we naively analyze the partial sky, the large
    spurious oscillations arise on the angular power spectrum.  In the
    lower panel, we show the cases with apodization, with $\theta_{\rm
    apd} = 100^\circ$.  The black and magenta lines are same as in the
    upper panel.  The red and blue lines represent the apodized $C_\ell$
    with Gaussian and Cosine types, respectively.  For both cases, the
    North West mask is used. The suppression of the spurious
    oscillations is clearly seen due to the apodization function.  }
    \label{fig:apornot}
\end{figure}

We calculate the angular power spectrum with apodization and show the
results in Fig.~\ref{fig:apornot}.  In the upper panel of the figure, in
the case of $\theta_{\rm max} =100^\circ$ when the North West mask
is used, the oscillatory feature with a large amplitude can be seen on
the angular power spectrum, because we truncate the integral in
Eq.(7) at $\theta=100^\circ$.  In the case of $\theta=180^\circ$
when KQ75 mask is used, on the other hand, the oscillation is
suppressed because $\xi$ decreases as $\theta \to 180^\circ$.  The lower
panel shows the results with apodization using the Gaussian and Cosine
type functions, setting $\theta_{\rm apd}$ to $100^\circ$.  As is shown in
the lower panel, the resultant power spectra become smooth compared to
the case without apodization (see the red line in the upper panel).
Also it is noted that the shape of the apodized spectra is similar to
the power spectrum of the full sky ($\theta_{\rm
max}=180^\circ$; the black line), though small scale oscillations have
smoothed out. This is a bad news because we are interested in the fine structures.

\begin{figure}[thpb]
  \begin{center}
    \includegraphics[width=0.35\textwidth, angle = -90]
    {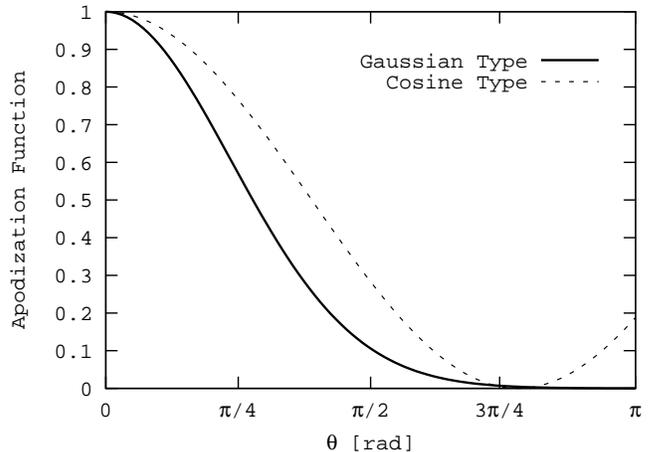}
  \end{center}
\caption{The apodization functions $F(\phi)$ of Cosine type (solid) and
 Gaussian type (dashed).  The $\theta_{\rm apd}$ parameters are fixed as
 they suppress enough the spurious oscillations ($\theta^{\rm cos}_{\rm
 apd}=140^\circ$, and $\theta_{\rm apd}^{\rm Gauss}=100^\circ$).  These lines indicate
 that the information lost is heavier for the case of Gaussian type
 apodization than that of Cosine type.}  \label{fig:apd_fun}
\end{figure}

%In our work, we use only the Cosine type apodization function, because
%we found that the Gaussian type function tends to lose the information
%about small scale feature compared to the Cosine one.  
As we take $\theta_{\rm apd}$ smaller, more significant the smoothing
effect for the fine structures in the angular power spectrum
becomes. Therefore we should take the angle $\theta_{\rm apd}$ as large
as possible to keep the fine structures.
The maximum apodization angles $\theta_{\rm apd}$ for our masks with a
sufficient suppression of the spurious oscillations are found to be
$140^\circ$ and $100^\circ$ for the Cosine and Gaussian types,
respectively.  In Fig.~\ref{fig:apd_fun} we depict the apodization
functions with those $\theta_{\rm apd}$.  From the figure, we can notice
that the Gaussian type reduces the information in the two-point correlation 
function at large angles more than the Cosine one.  
In this work, we should keep the information as much as possible,
because we focus on the fine structures.  Therefore, we use the
Cosine type apodization function in the following analysis.
In this case, the structure at $\ell = 100$--150 can survive the
apodization because 
the amplitude of the apodization function at this scale is over 0.99. 

As the loss of information about high frequency structure in the angular power spectrum
is significant, components of $C_\ell$
become correlated with each other. This effect and coping technique are
discussed in the next subsection.

\subsection{Binned angular power spectrum}

In the previous subsection, we see the apodization technique suppresses
the high frequency oscillations.  However, this technique is unavoidably
accompanied with the loss of information.  
The loss appears as correlations between two different multipole moments.  
In order to see the degree of 
the correlations, let us calculate the covariance matrix ${\cal
C}_{\ell\ell^{\prime}}$,
\begin{equation}
  \label{eq:cov_mat}
  {\cal C}_{\ell\ell^{\prime}}\equiv 
  \frac{1}{\bar C_{\ell}^2}\frac{1}{N}
  \sum_{i=1}^N \left(C_\ell^i - \bar C_\ell\right)
  \left(C_{\ell^\prime}^i - \bar C_{\ell^\prime}\right)~,
\end{equation}
where 
\begin{equation}
  \label{eq:clave}
  \bar C_\ell = \frac{1}{N}\sum_{i=1}^NC_\ell^i~.
\end{equation}
Here, $C_\ell^i$ is the estimated angular power spectrum from the $i$th
Monte Carlo simulation sky, N is the total number of the samples. From
this matrix, we can estimate the strength of correlations between each
multipole moment.  If there are no correlations, which means the
$C_\ell$ can be estimated independently, the covariance matrix should be
diagonal.

On the contrary, the covariance matrix has off-diagonal elements if the
$C_\ell$ depends on another multipole moment component $C_{\ell^\prime}$.
The correlations between the $C_\ell$'s are caused by the mask.  If a
mask is applied on the CMB sky, the estimated angular power spectrum is
given by a convolution between the power spectra of the mask and the
true temperature anisotropies that we want to estimate.  The convolution
generates correlations between the multipoles of $\Delta \ell \simeq$ 2--3,
which makes it complicated to estimate the statistical significance of
the angular power spectrum.  A simple way to obtain independent
observable is to bin the angular power spectrum with the comparable bin
width.

\begin{figure}[]
  \centering \includegraphics[width=0.5\textwidth]
  {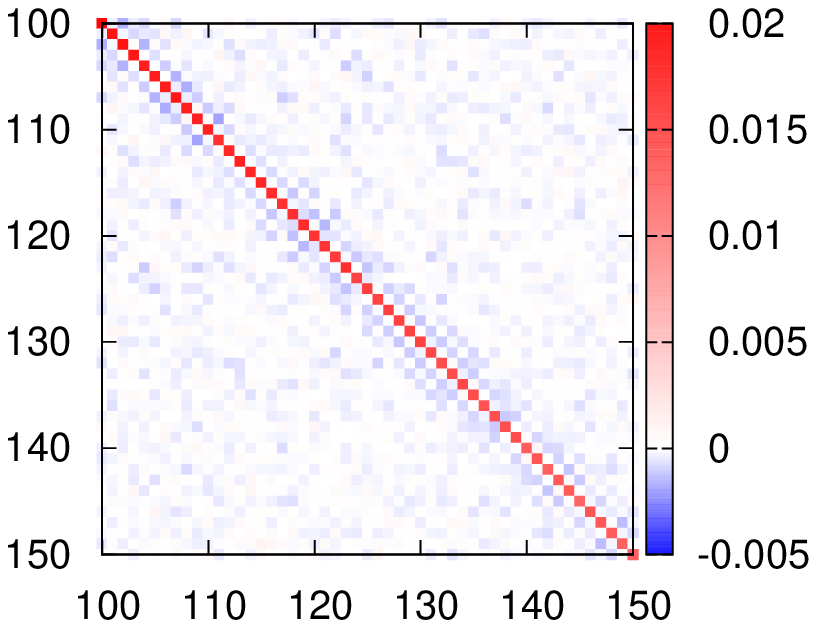} \centering
  \includegraphics[width=0.5\textwidth] {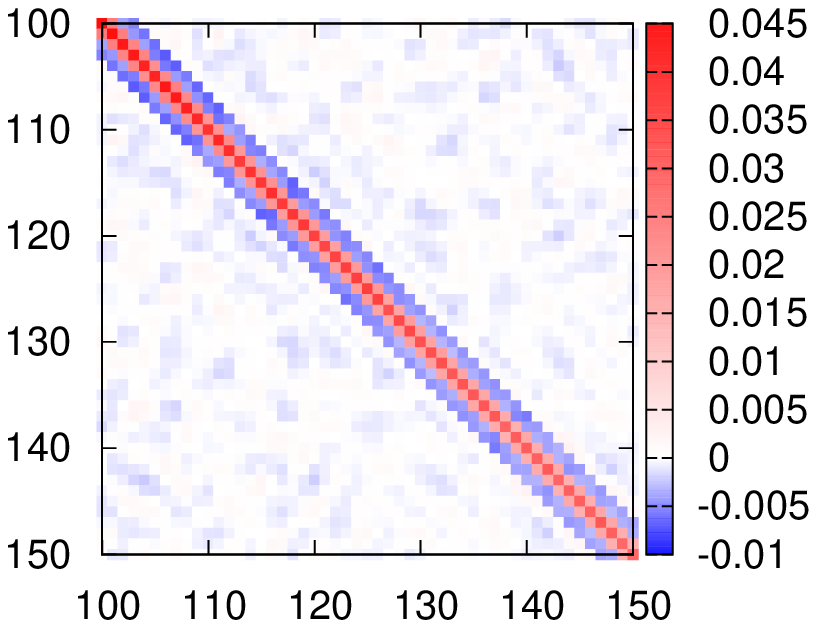}
  \centering \includegraphics[width=0.5\textwidth]
  {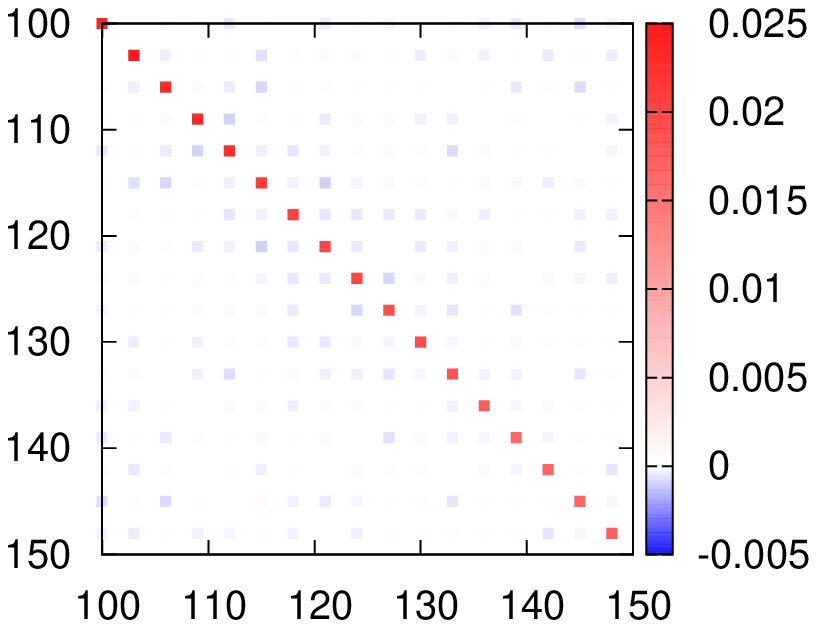} \caption{Covariance matrices of 
  angular power spectra ${\cal C}_{\ell\ell^{\prime}}$.  The upper and
  middle panels show the covariance matrices without apodization and
  with Cosine type apodization, respectively. Non-zero off-diagonal
  components indicate the presence of correlation between $\ell$ and
  $\ell^\prime$.  In the bottom panel we show the covariance matrix of the binned
  angular power spectrum.  We can see that the correlation between
  components becomes weak.  } \label{fig:cov}
\end{figure}

The covariance matrices of the angular power spectrum with KQ75 mask are
depicted in Fig.~\ref{fig:cov}.  The upper panel in the figure shows the
matrix for the case without apodization.  Each component of
this matrix is practically vanishing except for the diagonal ones.
This result indicates that the correlations caused by the KQ75 mask are
not significant at the multipole region of $100\leq \ell \leq 150$.
This is because the condition that $\Delta\ell \geq
\frac{\pi}{\theta_{\rm max}}\sim 1$ can still be satisfied with the
mask, where $\theta_{\rm max}$ is the maximum separation angle in the
pixel domain.  On the other hand, in the middle panel, we use the KQ75 mask
with apodization.  In this case, the components around diagonal elements
do not vanish.  This manifests the information loss due to apodization,
though the variances (diagonal elements) become smaller values.

The correlations appear only between the neighboring multipoles.  To
reduce the correlations we should gather the neighboring components in
one component, namely, bin the data.  The contributions from
the neighboring modes $\ell^\prime$ to the pivot scale $\ell$ are about 50\%
($\ell^\prime = \ell \pm 1$) and -10\% ($\ell^\prime = \ell \pm 2, \ell
\pm 3$), which is shown in the middle panel of Fig.~\ref{fig:cov}. We 
find that a binning with $\Delta \ell=3$ is sufficient because the
positive and negative correlations nicely compensate with each other to
reduce the correlations between the binned data.  In this case the
number of data points reduces to 17 from 51, and this reduction could
lead some information loss.  The covariance matrix of this case is
depicted in the lower panel of Fig.~\ref{fig:cov}.  Looking at this
figure, we can see that the correlation becomes weak enough thanks to
the binning.  Then, each components can be considered approximately
independent.

Although binning is useful to simplify the statistics, some information
should be lost in the process, especially if there exist fine structures
in the data.  In the next subsection we show how the binning procedure
affects the fine structures in multipole $\ell =100$--150 we are
interested in.  

\begin{figure*}[p]
  \begin{tabular}{cc}
    \begin{minipage}{0.6\textwidth}
      \centering
      \includegraphics[width=\textwidth]
      {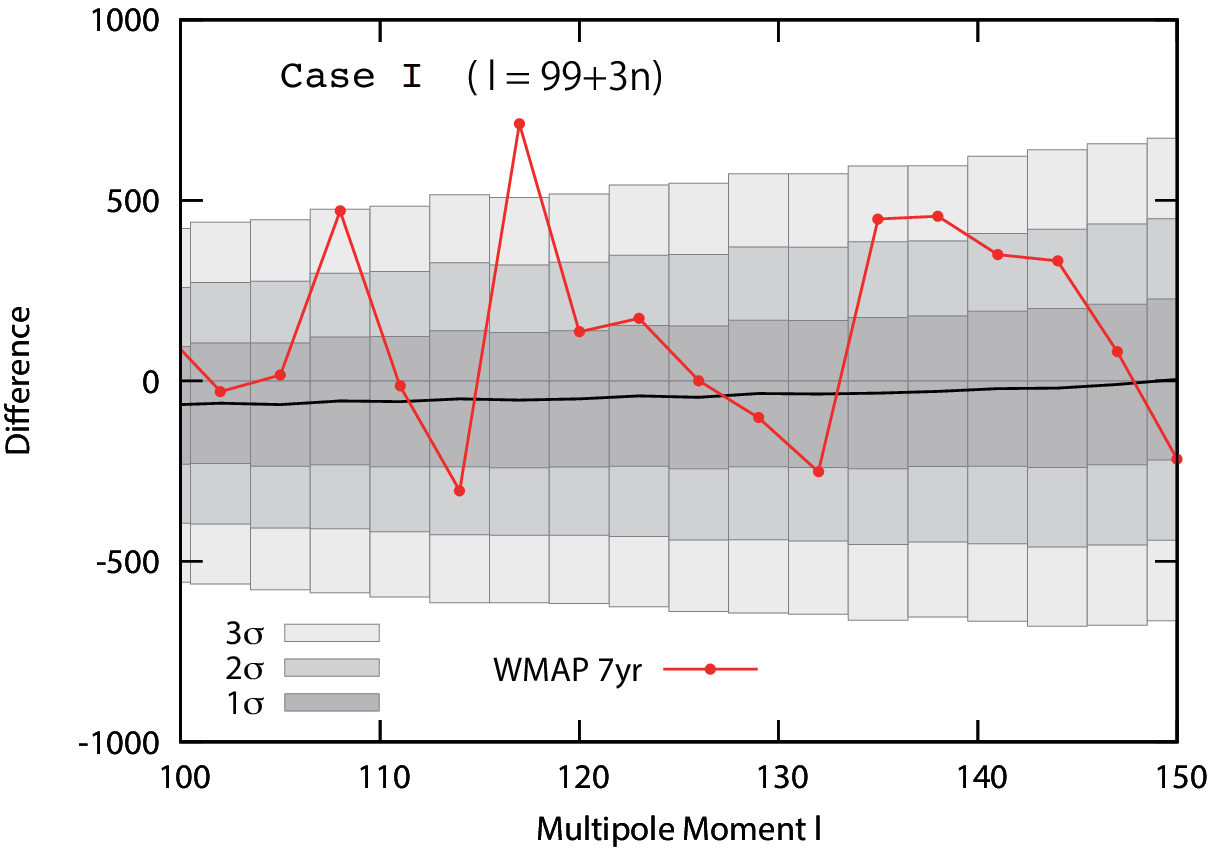}
    \end{minipage}
    \begin{minipage}{0.4\textwidth}
      \centering
        \includegraphics[width=0.9\textwidth]
        {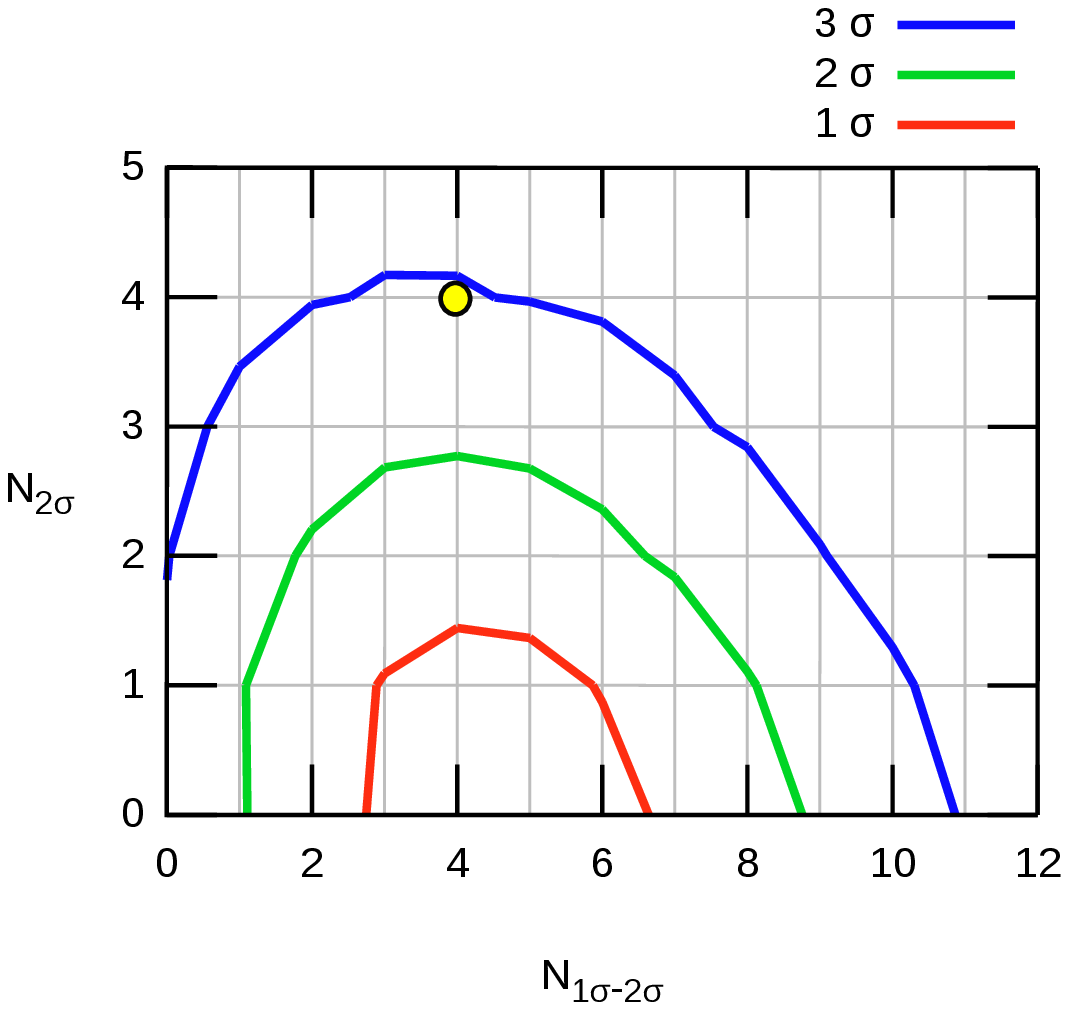}
    \end{minipage}
  \end{tabular}
  \begin{tabular}{cc}
    \begin{minipage}{0.6\textwidth}
      \centering
      \includegraphics[width=\textwidth]
      {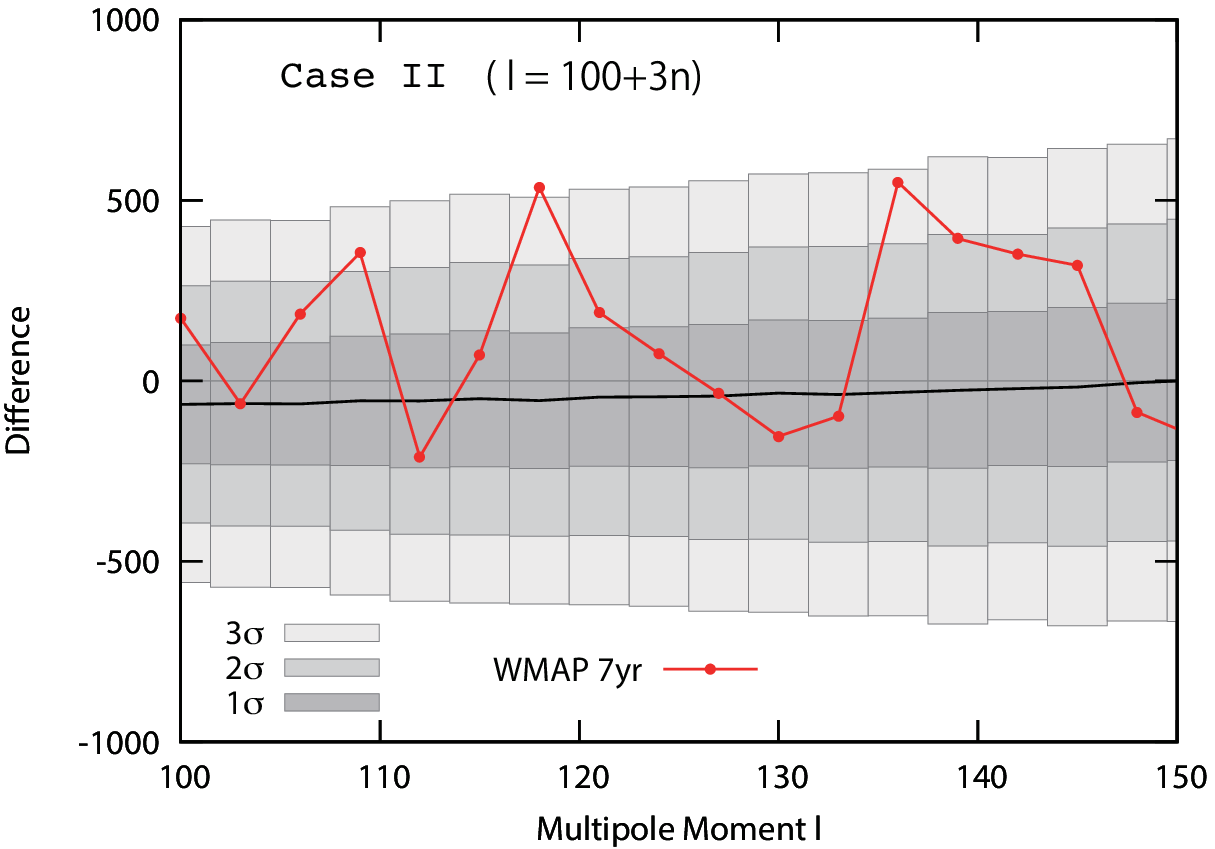}
    \end{minipage}
    \begin{minipage}{0.4\textwidth}
      \centering
        \includegraphics[width=0.9\textwidth]
        {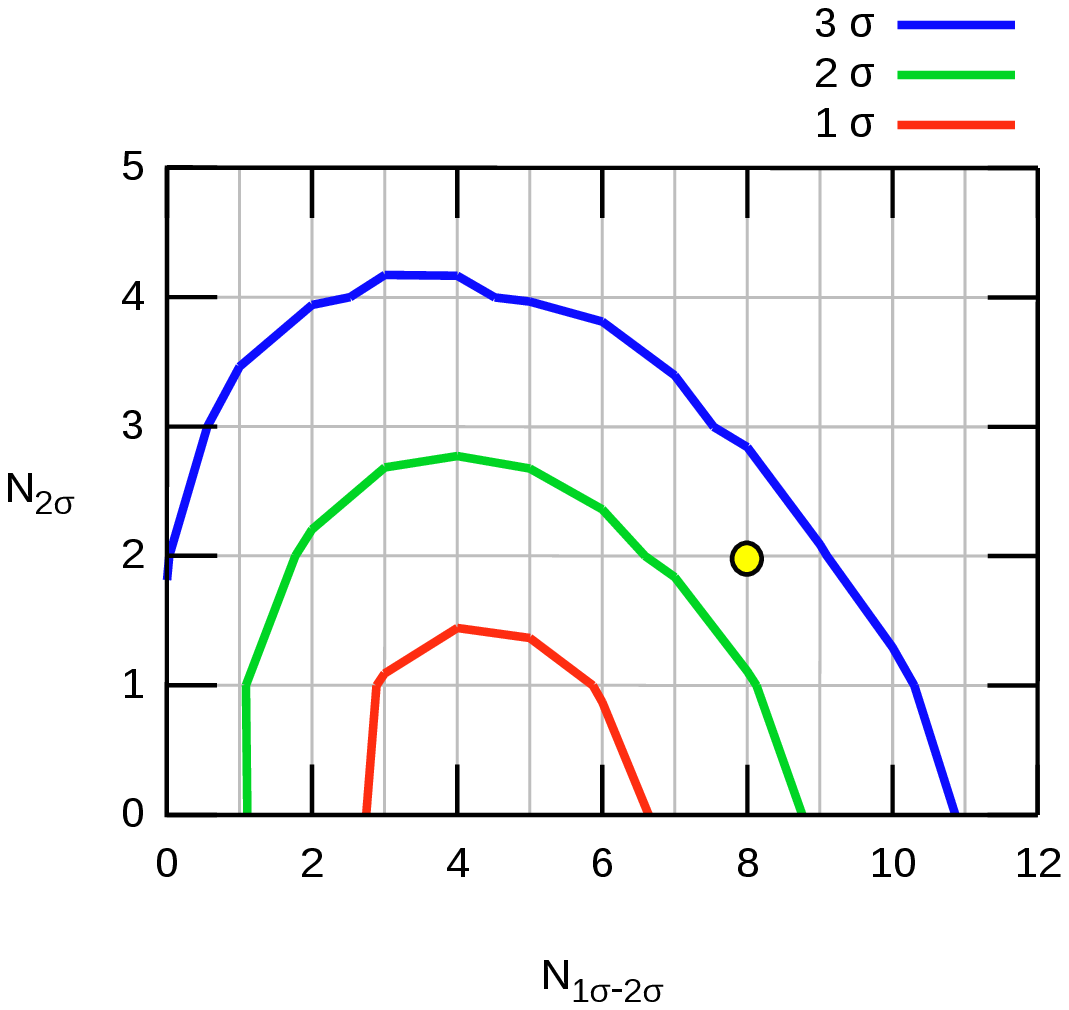}
    \end{minipage}
  \end{tabular}
  \begin{tabular}{cc}
    \begin{minipage}{0.6\textwidth}
      \centering
      \includegraphics[width=\textwidth]
      {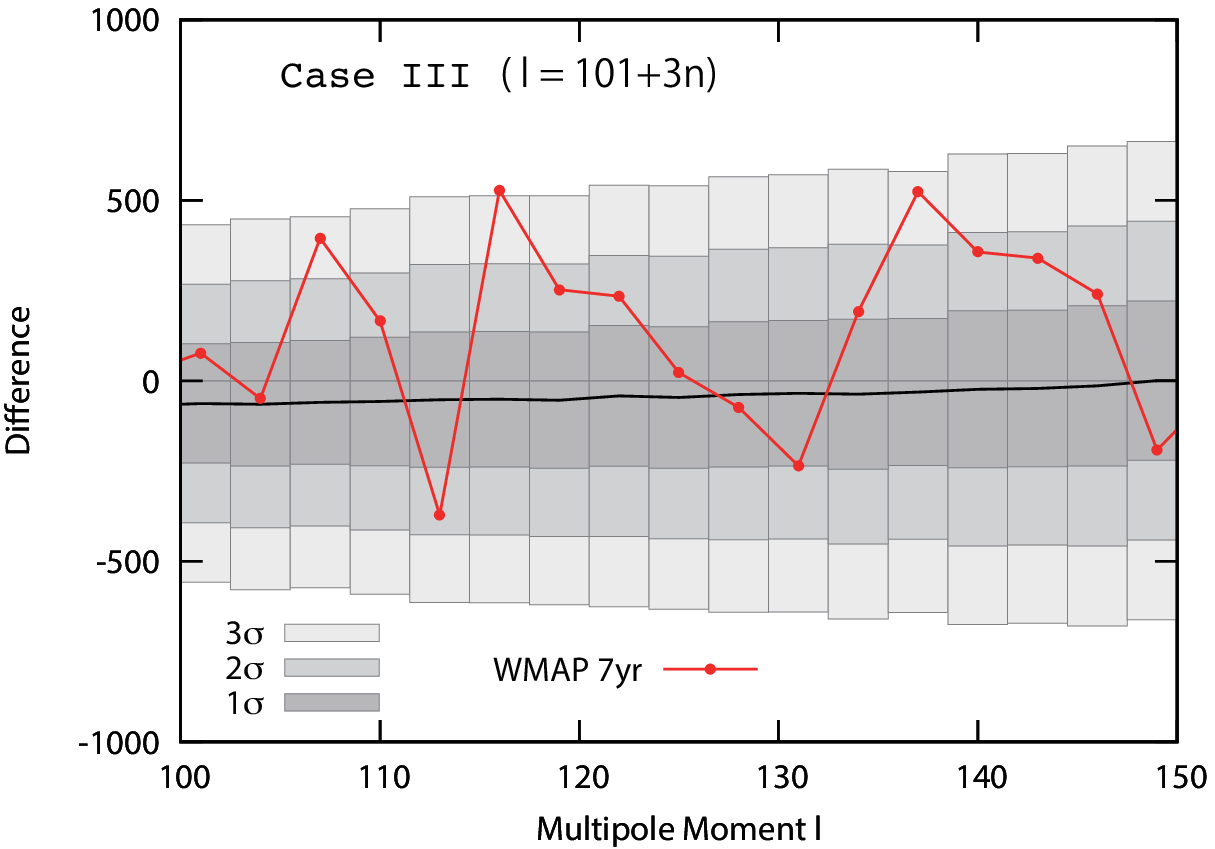}
    \end{minipage}
    \begin{minipage}{0.4\textwidth}
      \centering
        \includegraphics[width=0.9\textwidth]
        {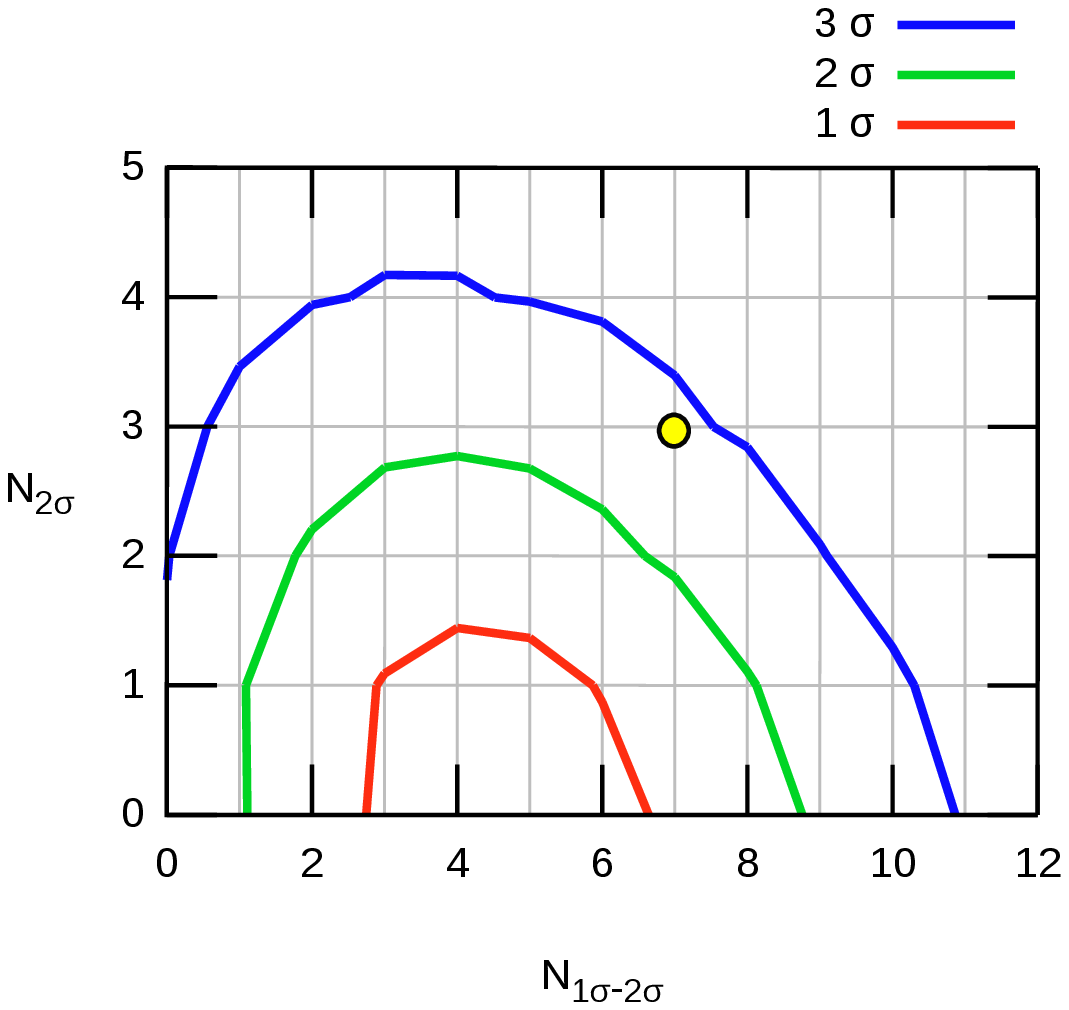}
    \end{minipage}
  \end{tabular}
  \caption{Left: Differences in the binned spectra between the predicted
 angular power spectrum from the $\Lambda$CDM model and the WMAP power
 spectrum with apodization function ($\theta_{\rm max} = 140^\circ$.
 The black line represents the average value of the 3,000 mock data.
 The deviation from zero manifests the $\chi^2$-like distribution of the
 angular power spectrum.  The boxes are the variances estimated from the
 mock data, as indicated in the figure.  Right: The probability contours
 assuming Gaussian distribution for each binned power with no
 correlation and the WMAP data, as in Fig.~2. }
 \label{fig:apdbincl}
\end{figure*}

We calculate the binned $C_\ell$'s and verify the fine structures using
the same steps we followed in Sec.~\ref{sec:vldty}.  There are three
ways to select the pivot scale for binning.  We calculate the binned
spectra with the three patterns $C_\ell^{\rm B} =
\sum_{i=-1}^1C_{\ell+i}/3$ for case I ($\ell = 99+3n$), case
I\hspace{-.1em}I ($\ell = 100+3n$), and case
I\hspace{-.1em}I\hspace{-.1em}I ($\ell = 101+3n$) where $n=1,2, \cdots,
17$.  The results are shown in Fig.~\ref{fig:apdbincl}.  The right
panels represent the probability that the fine structures come from the
noise and the cosmic variance effects, as we have shown in 
Fig.~\ref{fig:possibility}.
The result slightly depends on the binning pattern, but in any cases the
significance is above 2.5 $\sigma$.  In the left panels of
Fig.~\ref{fig:apdbincl}, we show the residuals between binned
$C_\ell^{\rm WMAP}$ and $C_\ell^{\rm \Lambda CDM}$ as red lines.  The
black line is the average of $C_\ell$ with the 3,000 simulations.  The
dispersion of the simulation data is also shown as the grey boxes.  We
can still see the oscillatory feature that is found in the case without
binning for all sky, and some points are over 1$\sigma$ box or 2$\sigma$
box.  The numbers of data points above 1 or 2 $\sigma$ are larger than
expected from Gaussian distribution for all cases.  These results
indicate that the fine structures can survive even if we apodize and bin
the angular power spectrum.

We consider these three power spectra as the fiducial binned power
spectra of WMAP for the full sky.  To figure out the origin of the fine
structures we compare them with the spectra from the partial skies
obtained with the same apodization function and binning, as we shall
show in the next section.  

\subsection{The angular power spectrum with partial sky}

In the previous sections we have seen that the apodization and binning
can successfully remove the spurious oscillations and the correlations
between multipoles due to the partial sky mask.  Here we apply the same
technique to the four different partial skies and see whether there
exists the directional dependence of the fine structures in the angular
power spectrum.  In the following discussion, we separate the structures
into two characteristic structures.  First one is the oscillatory
structure around $\ell = 100$ -- $120$.  The two peaks have a
significant deviation from the smooth spectrum at around 3$\sigma$.  The
other is the bump around $\ell = 140$.

\begin{figure*}[]
  \centering \includegraphics[width=0.9\textwidth]
  {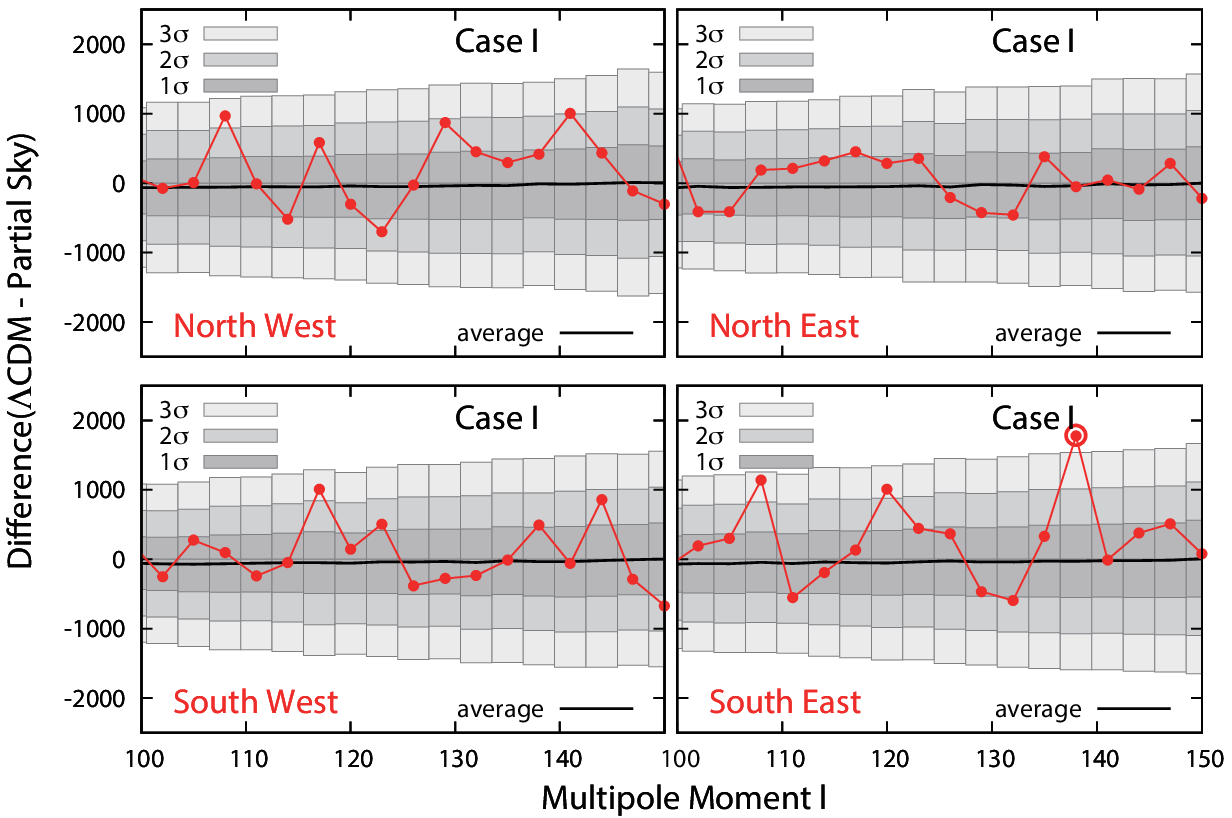} \includegraphics[width=0.9\textwidth]
  {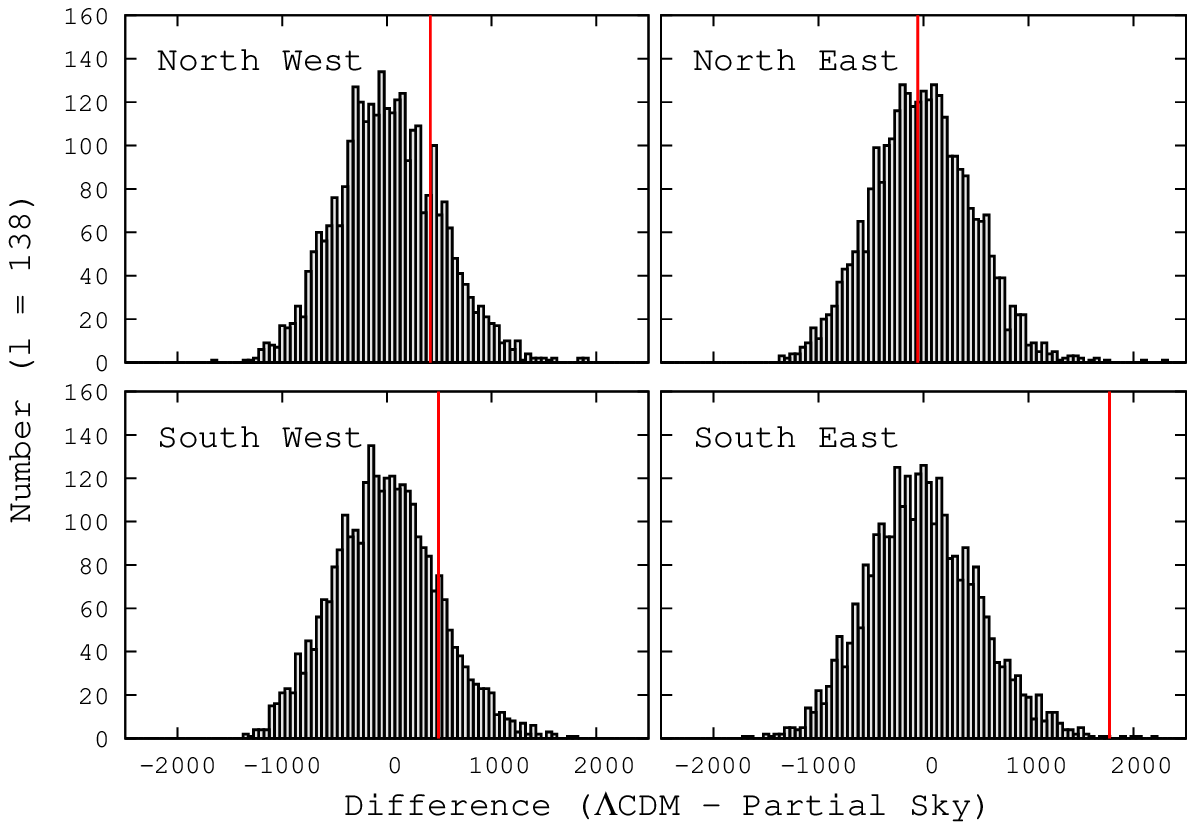} \caption{ Upper panels: The same as the left
  panels in Fig.~\ref{fig:apdbincl}, but for the partial skies with
  binning case I.  Lower panels: The histograms of differences between
  the power spectrum of $\Lambda$CDM and those from the 3,000 mock data
  at $\ell = 138$.  The red line corresponds to the data point of WMAP.
  The peculiarity of South East sky is clearly shown.} \label{fig:lcdm-areaz}
\end{figure*}

We show the results for the binning case I in the upper panel of
Fig.~\ref{fig:lcdm-areaz}.  The primary effect of the partial sky masks
can be seen in the error bars.  The estimated error bars depend on the
number of available modes $N$ as $1/\sqrt{N}$.  In our analysis the
effective sky coverage for the partial sky becomes about one-quarter
compared to the all sky map.  Therefore the error bar should be twice of
the case of the all sky (Fig.~\ref{fig:apdbincl}) and we can confirm this
fact from the figure.

\begin{figure*}[]
  \centering
  \includegraphics[width=0.9\textwidth]
  {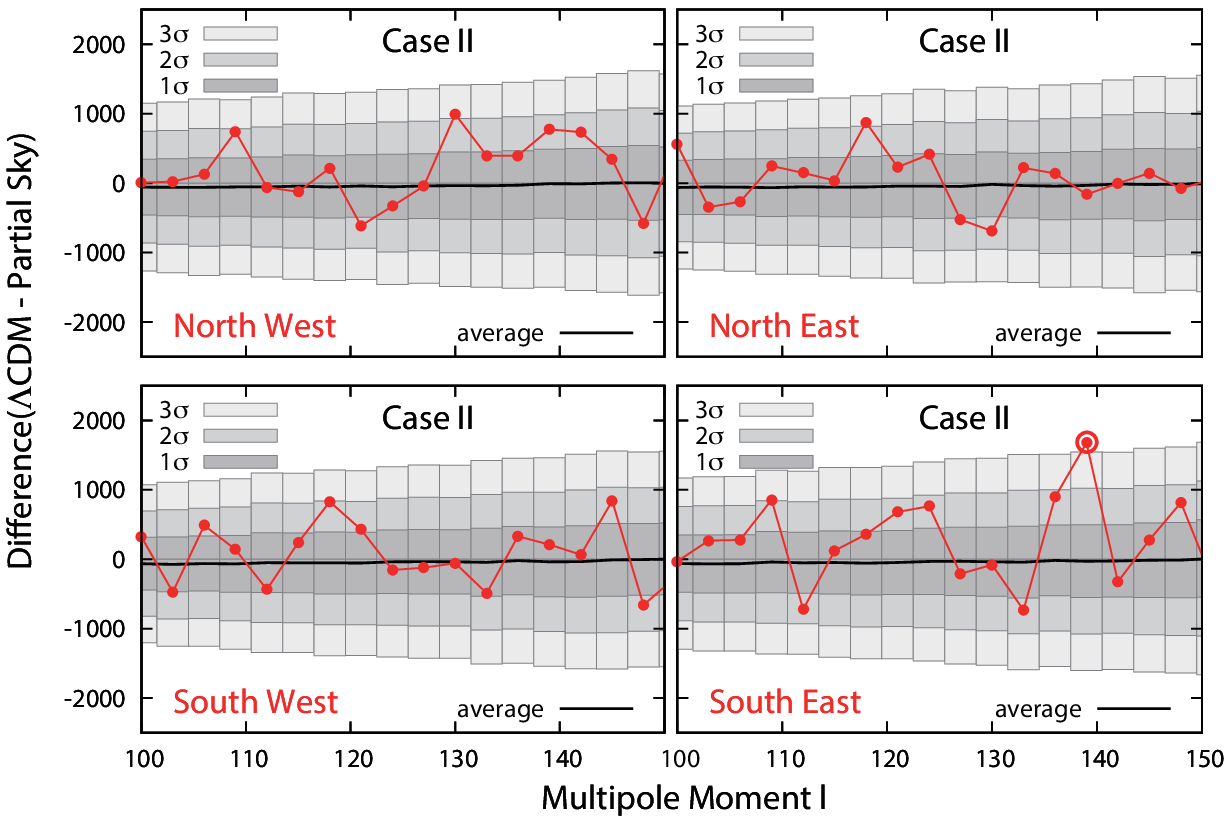}
  \includegraphics[width=0.9\textwidth]
  {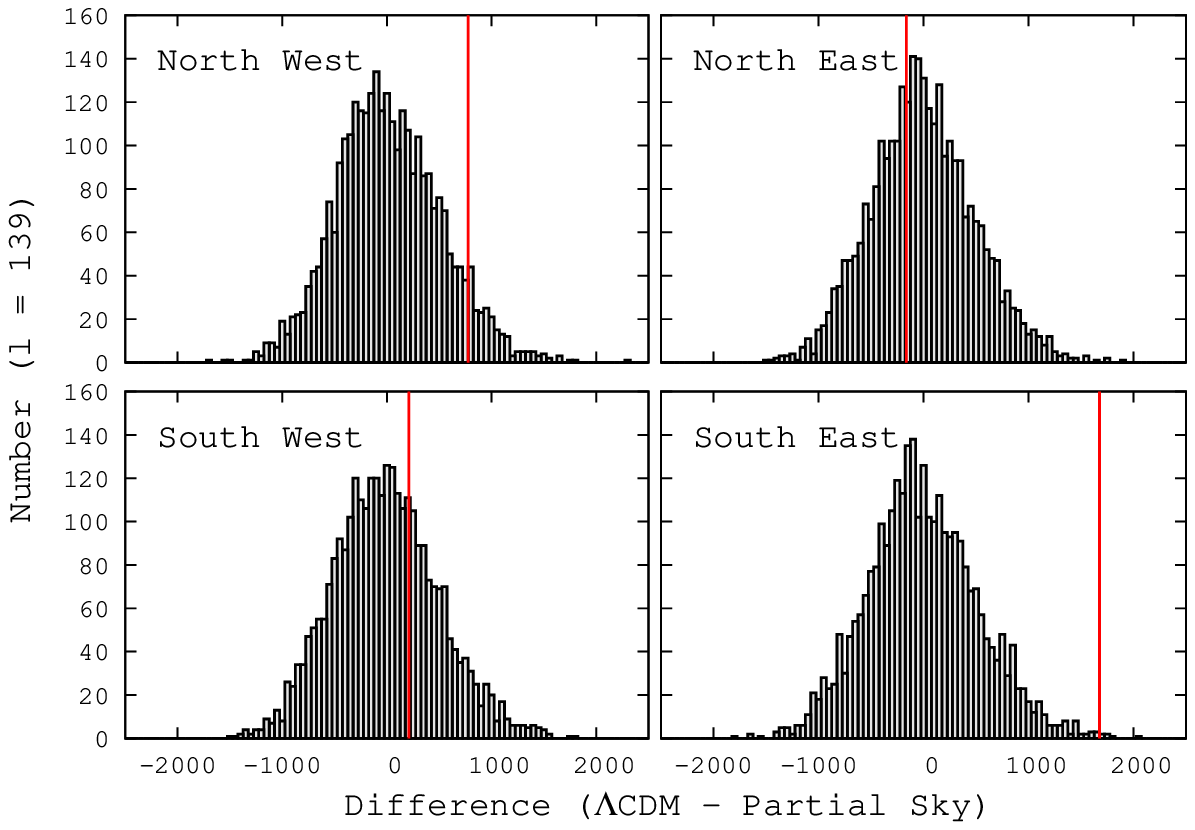}
  \caption{Same as Fig.~\ref{fig:lcdm-areaz} but for the binning case I\hspace{-.1em}I and 
    $\ell = 139$ (lower panel).  }
  \label{fig:lcdm-areao}
\end{figure*}

\begin{figure*}
  \centering
  \includegraphics[width=0.9\textwidth]
  {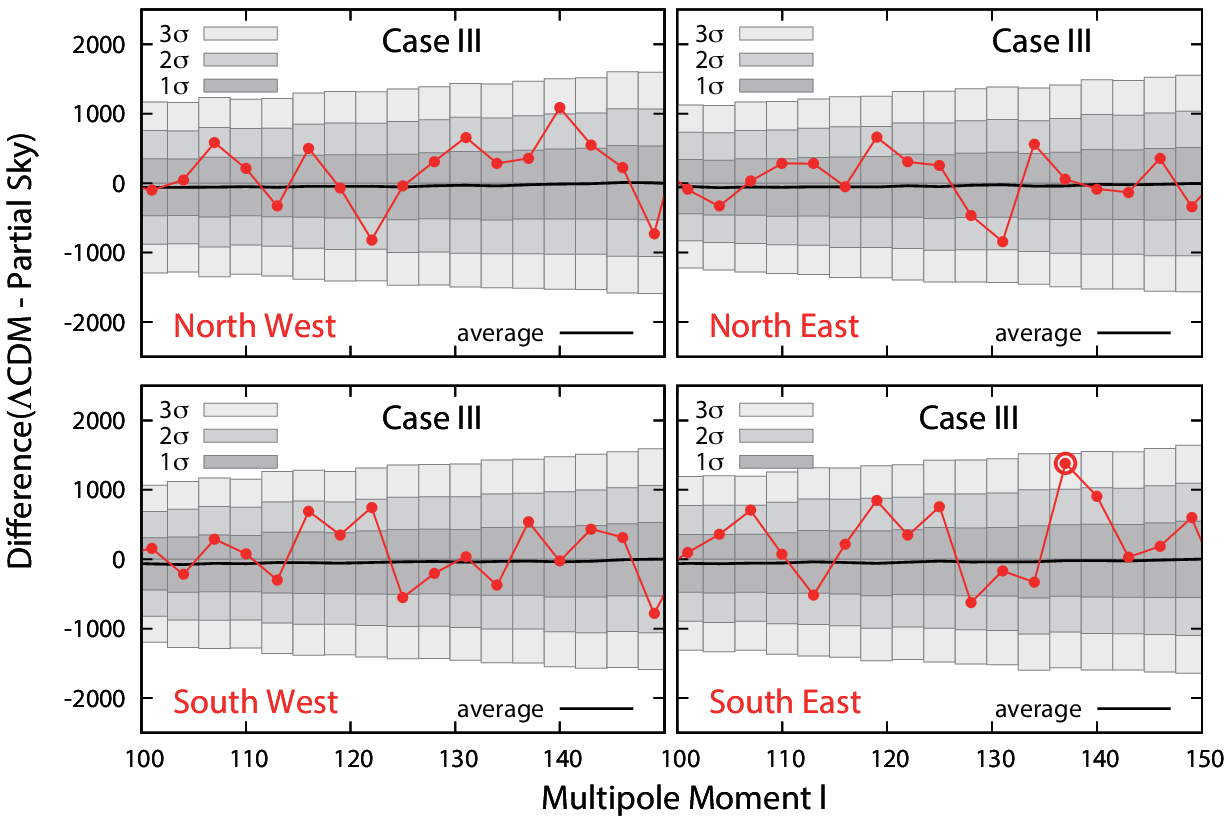}
  \includegraphics[width=0.9\textwidth]
  {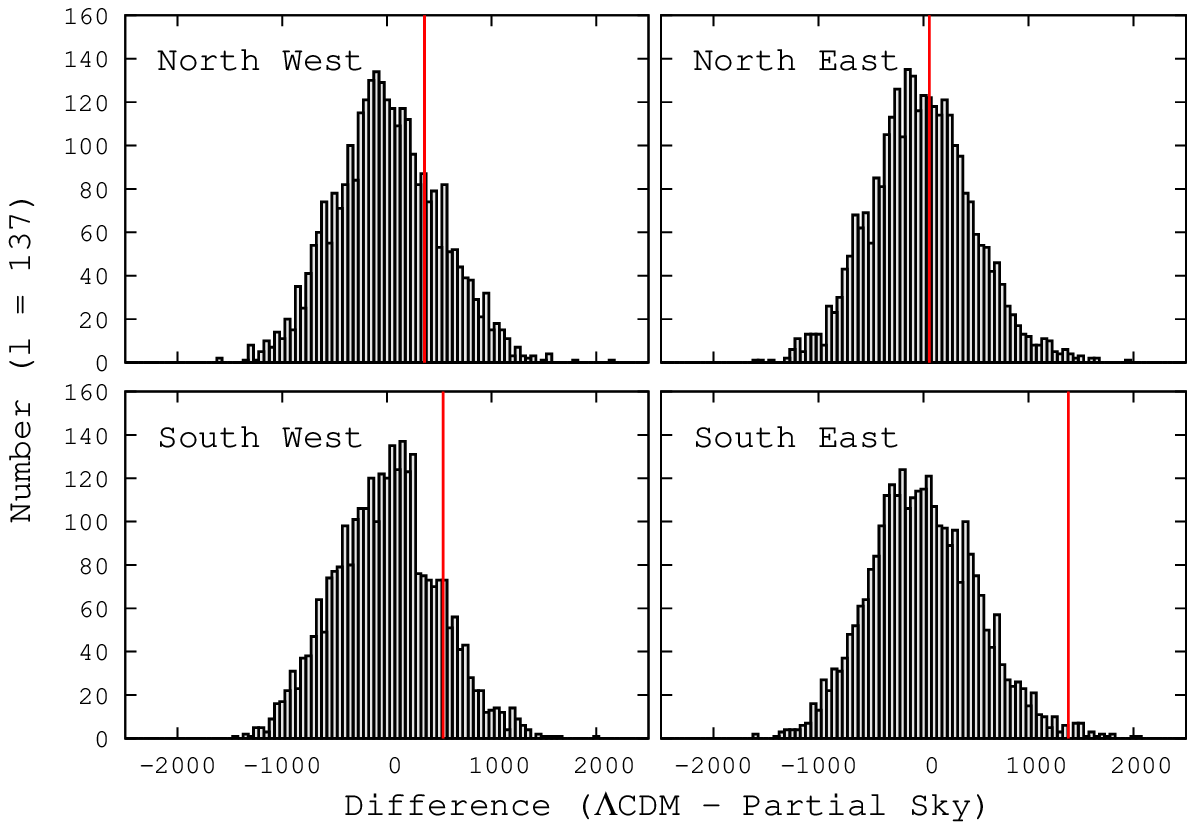}
  \caption{Same as Fig.~\ref{fig:lcdm-areaz} but for the binning case I\hspace{-.1em}I\hspace{-.1em}I
    and $\ell = 137$ (lower panel).  }
  \label{fig:lcdm-areat}
\end{figure*}

We find a distinct anisotropic structure which is above 3$\sigma$ error
bar at $\ell = 138$ in the South East area (bottom right panel of
Fig. \ref{fig:lcdm-areaz}), which is highlighted by a double circle
in figure.  Also, the histogram of the difference between
$\Lambda$CDM and the simulated data set at $\ell = 138$ is shown in the
lower panel of Fig.~\ref{fig:lcdm-areaz} for each direction.  The
vertical red lines show the differences of the power between the
$\Lambda$CDM and the observed one by WMAP.  The anisotropic structure in
the South East can be seen in the other binning cases, as in
Figs.~\ref{fig:lcdm-areao}, and \ref{fig:lcdm-areat}.

\begin{figure}[]
  \centering
  \includegraphics[width=0.5\textwidth]
  {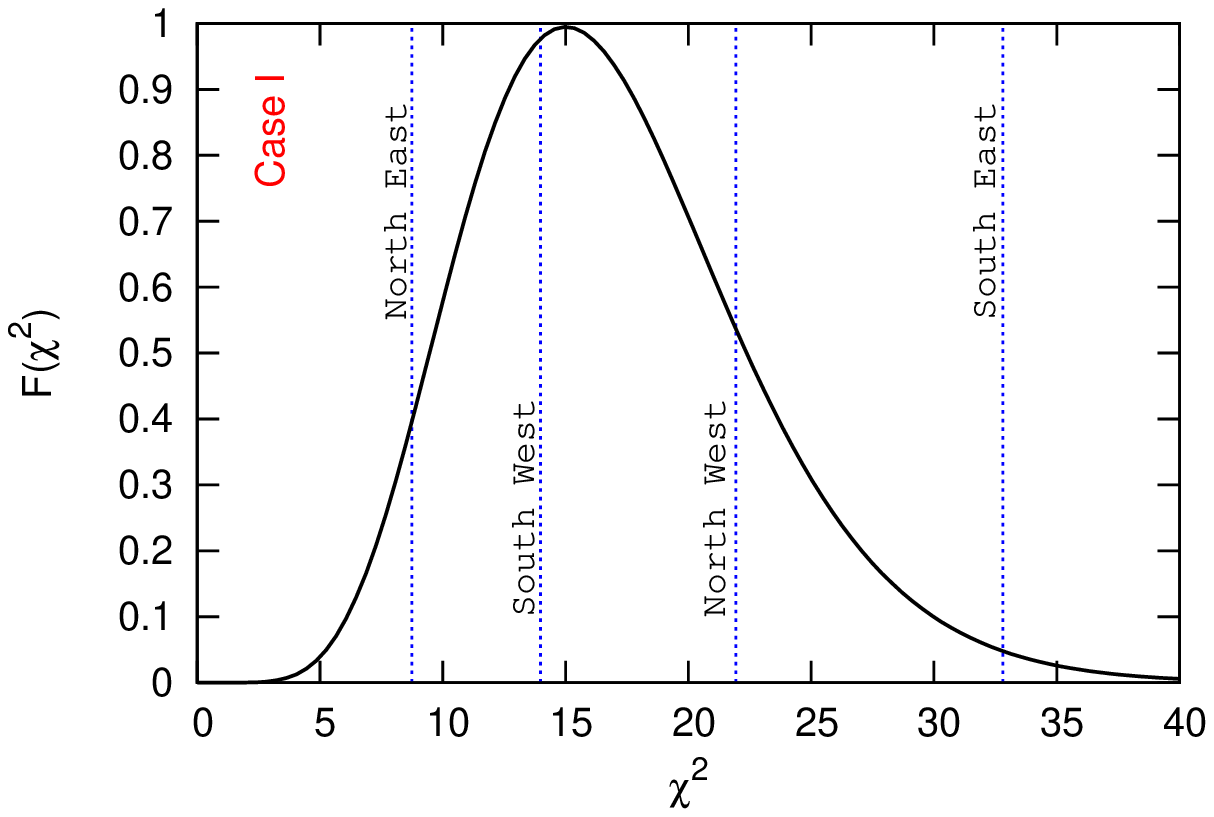}
  \centering
  \includegraphics[width=0.5\textwidth]
  {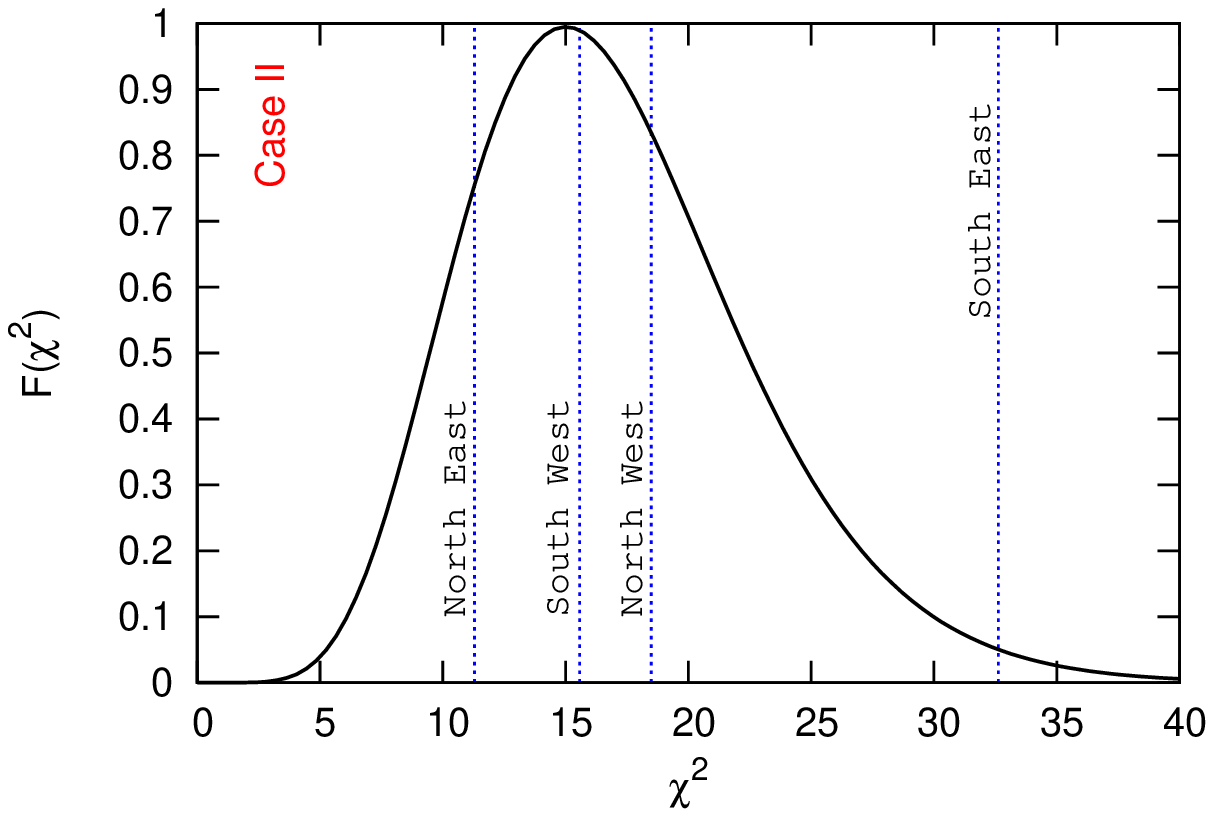}
  \centering
  \includegraphics[width=0.5\textwidth]
  {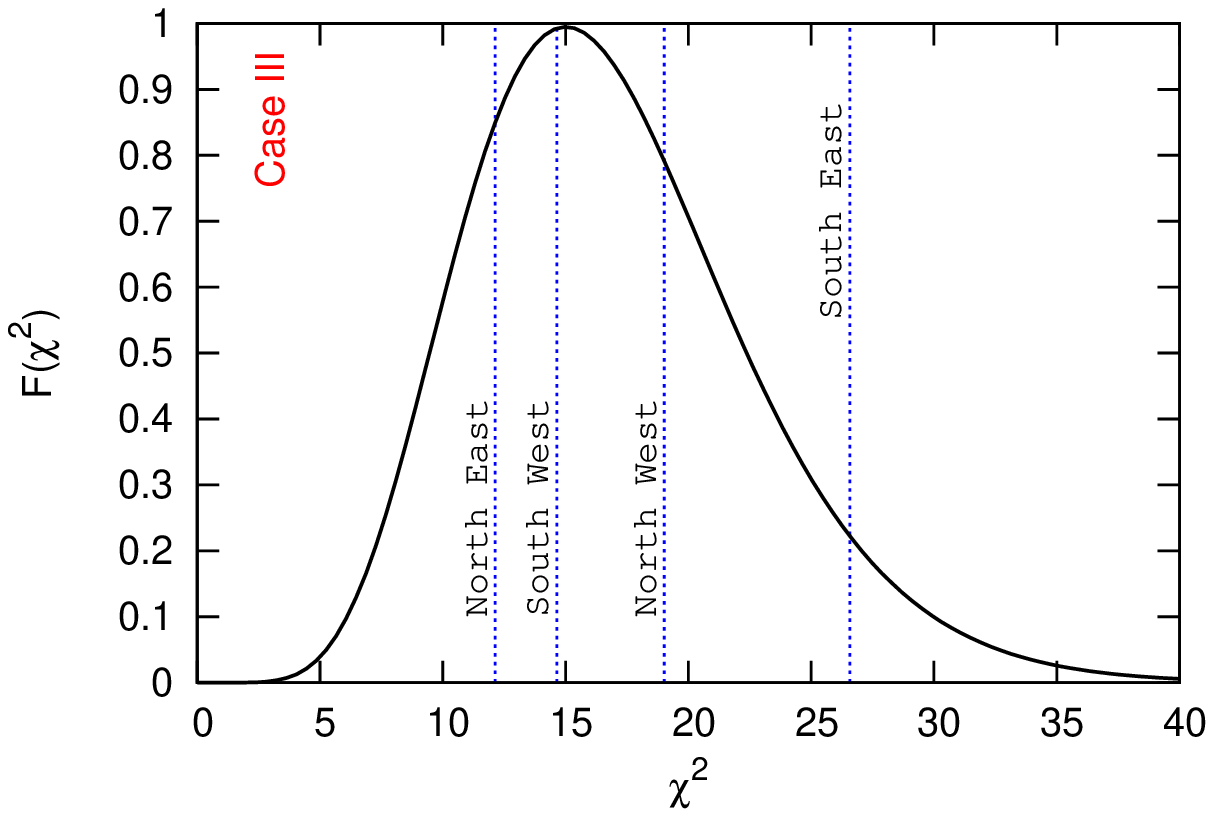}
  \caption{The $\chi^2$ distribution derived from the 3,000 simulations
 and the $\chi^2$ values of WMAP at each direction for each binning case.}
  \label{fig:chi2}
\end{figure}

The peculiarity of South East area as a whole at the multipole range of
$100 \leq \ell \leq 150$ can be quantified with the value of $\chi^2$,
defined by
\begin{equation}
\chi^2 = \sum_{i=1}^{17}\left(
C_\ell^{\rm B}-\bar{C}_\ell^{\rm B}
\right)
{\cal C}_{\ell\ell^\prime}^{{\rm B}^{-1}}
\left(
C_{\ell^\prime}^{\rm B}-\bar{C}_{\ell^\prime}^{\rm B}
\right)/\bar{C}^{\rm B}_{\ell} \bar{C}^{\rm B}_{\ell^\prime}~,
\end{equation}
where ${\cal C}_{\ell\ell^\prime}^{{\rm B}^{-1}}$ is the inverse of the
covariance matrix of the binned angular power spectrum.
We show the probability distribution function of $\chi^2$
, written by
\begin{equation}
F(\chi^2; n=17) = \frac{(\chi^2)^{n/2-1}}{2^{n/2}\Gamma(\frac{n}{2})}e^{-\chi^2/2}~,
\end{equation}
and the values of $\chi^2$
for the four partial skies in Fig.~\ref{fig:chi2}.  The value of
$\chi^2$ away from the peak position indicates the overall deviation of
the observed angular power spectrum from the average value of the mock
power spectrum.  From this figure, the South East area has especially
peculiar $\chi^2$ value.  The probability that the $\chi^2$ takes
larger value than the South East area by chance can be estimated as
1.50 (case I), 1.56 (case I\hspace{-.1em}I), 8.70 
(case I\hspace{-.1em}I\hspace{-.1em}I) \%.

These results may suggest a possibility of existence of characteristic
features only at South East area.  The structure which comes from
cosmological origin can be assumed to be isotropic, and therefore this
anisotropic structure around $\ell \approx 138$ might be attributed to
some astronomical origin, which have the scale about $0.6^\circ$. It
may be interesting to note that there has been a report about a power
spectrum anomaly around the third acoustic peak at the same sky
direction \cite{Ko:2011ut},
although the authors argued that the origin could be the WMAP
instrumental noise because the third acoustic peak is located at the
limit of the WMAP angular resolution.

The oscillatory feature (peak and dip) that is found in the all sky
analysis around $\ell = 100$ -- $120$, on the other hand, can be found
clearly in the North West area, and also found at all the other
directions regardless of the binning cases.  Thus, the oscillation seems
to be caused by some cosmological origin, not the astrophysical one.


%%%%%%%%%%%%%%%%%%%%%%%%%%%%%%%%%%%%%%%%%%%%%%%%
%%%
%%% Summary and discussion
%%%
%%%%%%%%%%%%%%%%%%%%%%%%%%%%%%%%%%%%%%%%%%%%%%%%

\section{Summary}

In this work we have re-examined the temperature angular power spectrum
of the cosmic microwave background (CMB) at three frequency bands and
for four restricted sky directions, in order to explore the origin of
the fine structures in the power at the multipole range of $100\leq \ell
\leq 150$.

We prepared 3,000 mock sky maps of CMB temperature fluctuations with
anisotropic instrumental WMAP noise, and verified the noise and cosmic
variance effect on the fine structures of the WMAP data.  By comparing
the angular power spectra from each mock data with the real data by
the WMAP, we found that the probability of the fine structures to be
realized by chance is about 2 - 3$\sigma$. We checked whether the fine
structures of the power spectrum depend on the Q, U, and V band maps,
and found no evidence for the frequency dependence.

In contrast, we obtained some interesting suggestions from the angular
power spectra derived from the partial skies.  We found that the
characteristic bump around $\ell=130$ -- $140$ seen in the all sky
angular power spectrum is solely attributed to the anomalous power at
the South East area. 
We have already known the existence of the cold spot as one of the 
anisotropy structures in South East area \cite{Cruz:2006fy, Inoue:2006rd}.
We have doubt if the cold spot will affect at $\ell =140$ as a substrucure,
thought the cold spot is few degree scale. 
However, we confirm that this effect is too tiny to generate characteristic structures
at $\ell = 140$. 
This result may indicate the existence of unknown peculiar
structure in that area as the origin of the fine structure.
The fine oscillating structures found around
$\ell = 100$ -- $120$, on the other hand, have no significant
directional dependences, suggesting that the oscillations come from some
cosmological origin.

If the observed feature is not a statistical fluctuation but has a
primordial origin, a straight but important test is to look into the
polarization of the CMB and/or the distribution of the large scale
structure, because they should show the same characteristic structure in
their power spectra. The coming data by PLANCK
\cite{2011A&A...536A...1P} and future galaxy survey such as the Large
Synoptic Survey Telescope \cite{2009arXiv0912.0201L} can be used for the
cross check and to accept (or reject) the feature with high
significance.

\begin{acknowledgements}
The authors would like to thank T. Matsumura, O. Tajima and M. Sato for helpful
 suggestions. This work is supported in part by the Grant-in-Aid for the
 Scientific Research Fund Nos. 24005235 (KK), 24340048 (KI) and 22340056 (NS)
 of the Ministry of Education, Sports, Science and Technology (MEXT) of
 Japan and also supported by Grant-in-Aid for the Global Center of
 Excellence program at Nagoya University "Quest for Fundamental
 Principles in the Universe: from Particles to the Solar System and the
 Cosmos" from the MEXT of Japan. This research has also been supported
 in part by World Premier International Research Center Initiative,
 MEXT, Japan.
\end{acknowledgements}

\bibliography{reference}
%\bibliography{mainNotes}

\end{document}